\documentclass[a4paper,11pt]{article}
\pdfoutput=1
\usepackage{amsmath}
\usepackage{amsfonts,amssymb}

\usepackage{jheppub}

\usepackage[T1]{fontenc}
\usepackage{bm}

\usepackage{graphicx}
\usepackage{lscape}
\usepackage{subfig}

\usepackage{mathtools}

\def\({\left(}
\def\){\right)}

\DeclareMathOperator{\dlog}{\mathit{d}log}

\def\eps{\epsilon}

\usepackage{xcolor}
\xdefinecolor{redviolet}{RGB}{228,38,207}
\xdefinecolor{brickred}{RGB}{220,60,66}
\xdefinecolor{bluscuro}{RGB}{51,51,179}

\title{\boldmath NNLO virtual and real leptonic corrections to muon-electron scattering}

\author[a,b]{Ettore Budassi,}
\author[b]{Carlo M. Carloni Calame,}
\author[a,b]{Mauro Chiesa,}
\author[a,b]{Clara Lavinia Del Pio,}
\author[b]{Syed Mehedi Hasan,}
\author[a,b]{Guido Montagna,}
\author[b]{Oreste Nicrosini }
\author[b]{and Fulvio Piccinini }

\affiliation[a]{Dipartimento di Fisica, Universit\`a di Pavia, Via A. Bassi 6, 27100 Pavia, Italy}
\affiliation[b]{INFN, Sezione di Pavia, Via A. Bassi 6, 27100 Pavia, Italy}

\emailAdd{ettore.budassi01@universitadipavia.it}
\emailAdd{carlo.carloni.calame@pv.infn.it}
\emailAdd{mauro.chiesa@unipv.it}
\emailAdd{claralavinia.delpio01@universitadipavia.it}
\emailAdd{syedmehe@pv.infn.it}
\emailAdd{guido.montagna@pv.infn.it}
\emailAdd{oreste.nicrosini@pv.infn.it}
\emailAdd{fulvio.piccinini@pv.infn.it}

\abstract{The recently proposed MUonE experiment at CERN aims at
  providing a novel determination of the leading order hadronic
  contribution to the muon anomalous magnetic moment through the study
  of elastic muon-electron scattering at relatively small momentum
  transfer. The anticipated accuracy of the order of 10ppm demands for
  high-precision predictions, including all the relevant radiative
  corrections.  The fixed-order NNLO radiative corrections due to the
  emission of virtual and real leptonic pairs are described and their
  numerical impact is discussed for typical event selections of the
  MUonE experiment, by means of the upgraded Monte Carlo code
  \textsc{Mesmer}.}
\begin{document}
\maketitle
\section{Introduction}
\label{sec:intro}
The value of the anomalous magnetic moment of the muon, $a_\mu =
(g-2)_\mu / 2$, is a fundamental quantity in particle physics, that
has been very recently measured by the Fermilab Muon $g-2$ Experiment
(E989) with the unprecedented relative precision of
$0.46$ppm~\cite{Abi:2021gix} for positive muons, improving on the
0.54ppm precision obtained by the E821 experiment at the Brookhaven
National Laboratory~\cite{Bennett:2006fi}.

The muon anomaly, {\it i.e.} the relative deviation of the
magnetic moment from the value predicted by Dirac theory, is due to quantum loop corrections
stemming from the QED, weak and strong sector of the
Standard Model (SM)~\cite{Jegerlehner:2009ry,Jegerlehner:2017gek}.
Hence, the comparison between theory and experiment provides a very stringent
test of the SM and a deviation from the SM expectation is a monitor for the detection of
possible New Physics signals.

There is a long-standing and puzzling muon $g-2$ discrepancy between
the measured value and the theoretical prediction, which presently
exceeds the $3\sigma$ level. In particular, the new E989 result is
$3.3\sigma$ greater than the theoretical prediction.  This excess
grows up to $4.2\sigma$ when the new experimental result is combined
with previous measurements using both $\mu^+$ and $\mu^-$
beams~\cite{Abi:2021gix}.  The current status of the SM theoretical
prediction for the muon $g-2$ has been recently reviewed in
ref.~\cite{Aoyama:2020ynm}.

Future runs of the E989 experiment at
Fermilab~\cite{Venanzoni:2014ixa,Grange:2015fou} and of the E34
experiment under development at
J-PARC~\cite{Iinuma:2011zz,Mibe:2011zz}, are expected to bring the
relative precision of the anomalous magnetic moment of the muon below
the level of $0.2$ppm.  This calls for a major effort on the theory
side in order to reduce the uncertainty in the SM prediction, which is
dominated by non-perturbative strong interaction effects as given by
the leading order hadronic correction, $a_\mu^\text{HLO}$, and the
hadronic light-by-light contribution. Actually, there is at present
general consensus that the latter has a small impact on the
theoretical uncertainty on $a_\mu$~\cite{Aoyama:2020ynm} (see also
refs.~\cite{Chao:2020kwq,Danilkin:2021icn,Chao:2021tvp} for very
recent investigations through dispersion relations and Lattice QCD
calculations).

Traditionally, $a_\mu^\text{HLO}$ has been computed via a dispersion
integral of the hadron production cross section in electron-positron
annihilation at low
energies~\cite{Jegerlehner:2018zrj,Keshavarzi:2018mgv,Davier:2019can,Aoyama:2020ynm}. More
recent discussions on issues and constraints on data used in
dispersion relations can be found in
refs.~\cite{Colangelo:2020lcg,Benayoun:2021ody}.  Lattice QCD
calculations are providing alternative evaluations of the leading
order hadronic
contribution~\cite{DellaMorte:2017dyu,Chakraborty:2017tqp,Borsanyi:2017zdw,Meyer:2018til,Blum:2018mom,Giusti:2018mdh,Giusti:2019xct,Giusti:2019hkz,Shintani:2019wai,Davies:2019efs,Gerardin:2019rua,Aubin:2019usy,Giusti:2020efo,Lehner:2020crt}. Lately
the BMW collaboration has presented a precise determination of
$a_\mu^\text{HLO}$ with an uncertainty of
$0.78\%$~\cite{Borsanyi:2020mff}, a central value larger than the ones
obtained via dispersive approaches, by an amount that ranges from $2$
to $2.5\sigma$ depending on the reference value used for the
dispersive approach, and in agreement with the BNL and FNAL
experimental determinations~\cite{Bennett:2006fi,Abi:2021gix}. To
clarify this tension, alternative and independent methods for the
evaluation of $a_\mu^\text{HLO}$ are therefore more than welcome and
desirable.

Recently, a novel approach has been proposed in ref.~\cite{Calame:2015fva}
to derive $a_\mu^\text{HLO}$ from a measurement of the effective
electromagnetic coupling constant in the space-like region via
scattering data, making use of a relation between
$\Delta\alpha_\text{had}(q^2)$ at negative squared momenta and
$a_\mu^\text{HLO}$ (see also~\cite{Lautrup:1971jf}).
Shortly afterwards, the elastic scattering of high-energy muons on atomic electrons has been
identified as an ideal process for such a
measurement~\cite{Abbiendi:2016xup}~\footnote{A method to measure the
running of the QED coupling in the space-like region using small-angle
Bhabha scattering was proposed in ref.~\cite{Arbuzov:2004wp} and
applied to LEP data by the OPAL Collaboration~\cite{Abbiendi:2005rx}.
In the time-like region, the effective QED coupling constant in the
region below $1\text{ GeV}$ has been recently measured by the KLOE collaboration~\cite{KLOE-2:2016mgi}.
}
and a new experiment, MUonE, has been proposed at CERN to measure the
differential cross section of $\mu e$ scattering as a function of the
space-like squared momentum
transfer~\cite{MUonE:LoI,Abbiendi:2019qtw,Ballerini:2019zkk,Abbiendi:2021xsh}.
In order for this new determination of $a_\mu^\text{HLO}$ to be
competitive with the traditional dispersive approach, the uncertainty
in the measurement of the $\mu e$ differential cross section must be
of the order of 10ppm.

The MUonE challenging experimental target requires very high-precision
predictions for $\mu e$ scattering, including all the relevant
radiative corrections, implemented in fully-fledged Monte Carlo (MC)
tools. The necessary perturbative accuracy will be next-to-next-to
leading order in QED (NNLO QED), combined with the leading (and,
eventually, next-to-leading) logarithmic contributions due to multiple
photon radiation. In recent years, a number of steps have been already
taken to achieve this goal. A comprehensive review of the theoretical
knowledge, as of the beginning of year 2020, of the $\mu e$ scattering
cross section for MUonE kinematical conditions has been published in
ref.~\cite{Banerjee:2020tdt}.  We list below the main milestones
already reached in this theoretical endeavour.  In
ref.~\cite{Alacevich:2018vez}, the full set of NLO QED and one-loop
weak corrections was computed without any approximation and
implemented in a fully exclusive MC generator. It is presently being
used for simulation studies of MUonE events in the presence of QED
radiation. Important results were also obtained at NNLO accuracy in
QED. The master integrals for the two-loop planar and non-planar
four-point Feynman diagrams were computed in
refs.~\cite{Mastrolia:2017pfy,DiVita:2018nnh}, by setting the electron
mass to zero while retaining full dependence on the muon mass.  A
general procedure to extract leading electron mass terms for processes
with large masses, such as muon-electron scattering, from the
corresponding massless amplitude was given in
ref.~\cite{Engel:2018fsb} and supplemented with a subtraction scheme
for QED calculations with massive fermions at NNLO
accuracy~\cite{Engel:2019nfw}.  More recently, the exact NNLO photonic
corrections along the electron line, including all finite mass terms,
have been implemented in two independent MC tools,
\textsc{Mesmer}~\cite{CarloniCalame:2020yoz} and
\textsc{McMule}~\cite{Banerjee:2020rww}.  In the former, the framework
for the complete photonic NNLO calculation has been built and made
available, including the full NLO calculation to the process $\mu
e\to\mu e\gamma$ and the LO double radiative process $\mu e\to \mu
e\gamma\gamma$.  Since the two-loop diagrams where at least two
virtual photons connect the electron and muon lines are not completely
known yet, their infrared (IR) part has been taken into account by
means of the classical Yennie-Frautschi-Suura
approach~\cite{Yennie:1961ad}.  Very recently, the analytic evaluation
of the two-loop corrections to the amplitude for the scattering of
four fermions in QED has been carried out keeping the complete
dependence on the mass of one fermionic
current~\cite{Bonciani:2021okt}.  The two-loop hadronic corrections to
$\mu e$ scattering were computed in
refs.~\cite{Fael:2018dmz,Fael:2019nsf}. Also possible contamination
from New Physics effects has been studied in
refs.~\cite{Dev:2020drf,Masiero:2020vxk} and shown to be below the
MUonE sensitivity, thus reinforcing the robustness of the proposed
approach for a reliable determination of $a_\mu^{\rm HLO}$ through
MUonE data.

In this paper, we present the calculation of the complete fixed-order
NNLO QED corrections which include at least one leptonic pair, with
all the relevant virtual and real contributions, summing over all
contributing lepton flavours. This class of corrections is named NNLO
leptonic correction.  In section~\ref{sec:class} we describe the
classification of the contributions and the methods used for their
calculation. Where needed, the calculation of the virtual corrections
is performed by means of the dispersion relation
technique~\footnote{The same approach has been used in the literature
for the calculation of two-loop QED hadronic and leptonic corrections
to Bhabha
scattering~\cite{Actis:2007fs,Actis:2008sk,Kuhn:2008zs,CarloniCalame:2011zq}.}
already used for the hadronic NNLO corrections in
refs.~\cite{Fael:2018dmz,Fael:2019nsf}, taking into account all finite
mass effects.  The real-pair emission contributions $\mu^\pm
e^-\to\mu^\pm e^- e^+ e^-$ and $\mu^\pm e^-\to\mu^\pm e^- \mu^+ \mu^-$
are calculated by means of exact tree-level matrix elements, including
electron and muon finite mass effects. For completeness, we calculate
also the NNLO virtual hadronic corrections, already calculated in
refs.~\cite{Fael:2018dmz,Fael:2019nsf}, and compare their size with
the leptonic corrections on all relevant differential distributions.
The whole class of corrections has been included in an upgraded
version of the \textsc{Mesmer}~\footnote{The code is now available on
the public repository \href{https://github.com/cm-cc/mesmer/}{\tt
  github.com/cm-cc/mesmer}.} MC code. By means of the
developed MC code we illustrate in section~\ref{sec:numerics}
numerical results relevant for typical running conditions and event
selections of the MUonE experiment.

The effects induced by the insertion of a leptonic loop are discussed
in sections~\ref{sec:numerics-factorized-on-born},~\ref{sec:numerics-vp-on-vertex}
and~\ref{sec:numerics-vp-on-photRC} and the impact of real lepton-pair
emission is discussed in
sections~\ref{sec:numerics-real-pairs-inclusive}
and~\ref{sec:numerics-real-pairs-exclusive}, with attention to the
interplay between real and virtual pair radiation as well as to the
role played by the ``peripheral'' diagrams within realistic MUonE
event selection.

A summary and future prospects are finally given in section~\ref{sec:summary}.
The work presented in this paper represents an additional step
towards the implementation of a fully-fledged MC generator including the
complete set of NNLO corrections matched to multiple photon emission.

\section{Classification of NNLO leptonic contributions}
\label{sec:class}
The complete set of NNLO leptonic corrections to $\mu^\pm e^-\to\mu^\pm e^- $ scattering, which
we denote with $d\sigma_{N_f}^{\alpha^2}$, consists of
three parts, with contributions from virtual and real leptonic corrections,
which can be schematically indicated with the following equation
\begin{equation}
  d\sigma_{N_f}^{\alpha^2}= d\sigma_\text{virt}^{\alpha^2}
  + d\sigma_\gamma^{\alpha^2} + d\sigma_\text{real}^{\alpha^2}\, .
  \label{eq:class-def}
\end{equation}

For later use, we remind that the running of the QED coupling constant
$\alpha$ through Dyson resummation of 1PI diagrams, in the on-shell
renormalisation scheme which is adopted here, reads
\begin{equation}
\alpha(q^2) = \frac{\alpha}{1-\Delta\alpha(q^2)}
    \label{eq:QEDalpharun}
\end{equation}
where $\Delta\alpha(q^2)$ can be expanded in $\alpha$ as
\begin{equation}
    \Delta\alpha(q^2) = \sum_{i=e,\mu,\tau,\text{top}}\left[\Delta\alpha^\text{LO}_{i}(q^2) + \Delta\alpha^\text{NLO}_{i}(q^2)\right] + \Delta\alpha^\text{hadr}(q^2)\, .
\label{eq:Deltaalpha}
\end{equation}
In the previous equation the sum of the perturbative bits runs over
all leptons and top quark, while the non-perturbative part
$\Delta\alpha^\text{hadr}(q^2)$, due to the exchange of hadronic
states, is traditionally evaluated through dispersion relations and
the optical theorem. We remark that in QED including the running of
$\alpha$ in a scattering amplitude amounts to replace the photon
propagator of virtuality $q^2$, multiplied by the vertex couplings,
according to
\begin{equation}
    \frac{e^2}{q^2}\to \frac{e^2}{q^2}\; \frac{1}{1-\Delta\alpha(q^2)}\, .
    \label{eq:propreplacement}
\end{equation}
\begin{figure}[t]
\begin{center}
\includegraphics[width=0.25\textwidth]{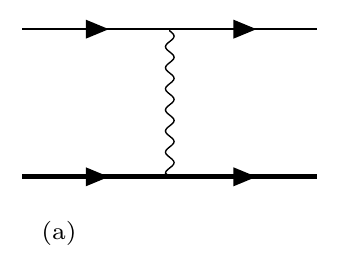}\hspace{0.5cm}
\includegraphics[width=0.25\textwidth]{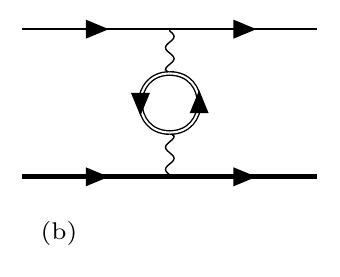}
\caption{LO diagram, (a), and NLO diagram, (b), 
with leptonic corrections. 
The thick fermionic line represents the muon 
current, while the double line closed loop stands
for the virtual leptonic contribution due to a generic lepton $\ell$.}
\label{Fig:lepton-one-loop}
\end{center}
\end{figure}
\begin{figure}[t]
\begin{center}
\includegraphics[width=0.23\textwidth]{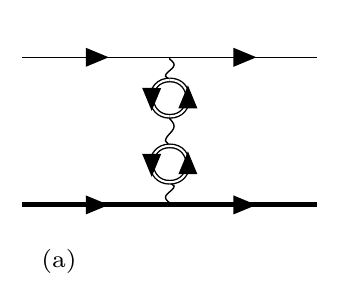}\hspace{0.5cm}
\includegraphics[width=0.23\textwidth]{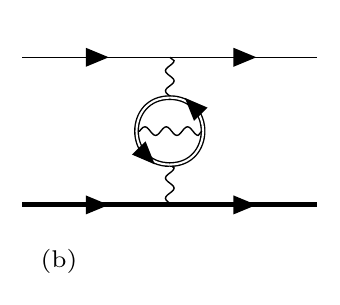}
\caption{
Left: NNLO reducible photon vacuum polarisation contribution.
Right: QED NNLO irreducible vacuum polarisation contribution.
}
\label{Fig:vacpol2}
\end{center}
\end{figure}
\begin{figure}[t]
\begin{center}
\includegraphics[width=0.23\textwidth]{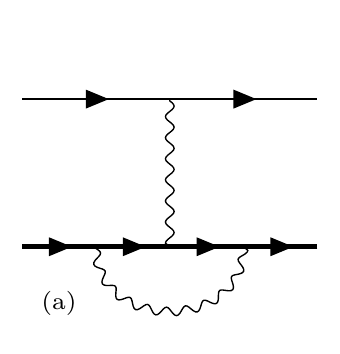}\hspace{0.1cm}
\includegraphics[width=0.23\textwidth]{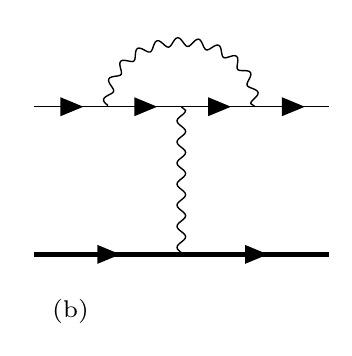}\hspace{0.1cm}
\includegraphics[width=0.23\textwidth]{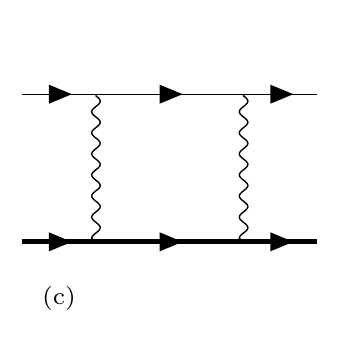}\hspace{0.1cm}
\includegraphics[width=0.23\textwidth]{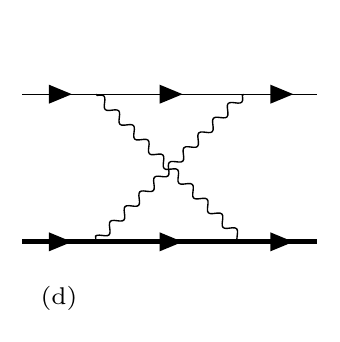}
\caption{NLO virtual photonic corrections.}
\label{Fig:one-loop-phot}
\end{center}
\end{figure}
\begin{figure}[t]
\begin{center}
\includegraphics[width=0.23\textwidth]{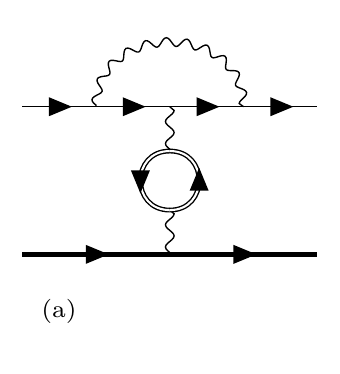}\hspace{0.5cm}
\includegraphics[width=0.23\textwidth]{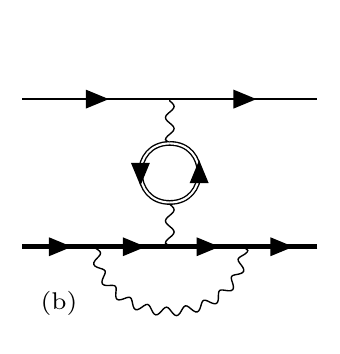}
\caption{Factorisable NNLO contribution with vacuum polarisation insertion on the tree-level photon propagator of one-loop photonic correction diagrams.
        }
        \label{Fig:red-vert} 
\end{center}
\end{figure}
\begin{figure}[t]
        \begin{center}
          \includegraphics[width=0.23\textwidth]{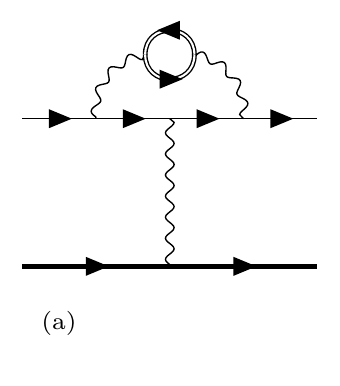}\hspace{0.5cm}
          \includegraphics[width=0.23\textwidth]{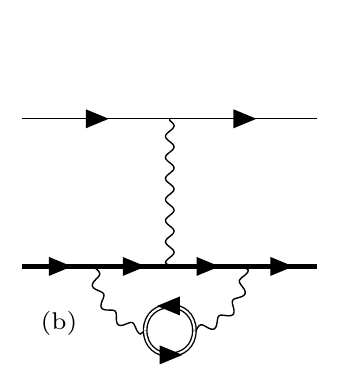}
        \caption{
          NNLO irreducible vertex diagrams. Vertex correction
          along the electron line (left) and along the muon line (right).}
\label{Fig:irred-vert}
\end{center}
\end{figure}
\begin{figure}[t]
        \begin{center}
          \includegraphics[width=0.23\textwidth]{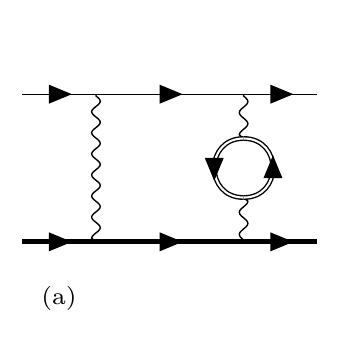}\hspace{0.1cm}
          \includegraphics[width=0.23\textwidth]{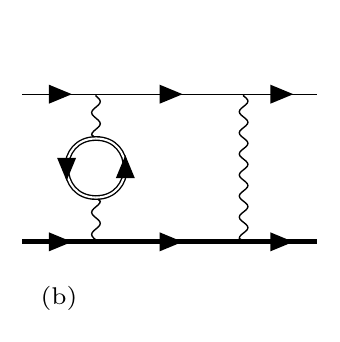}\hspace{0.1cm}
          \includegraphics[width=0.23\textwidth]{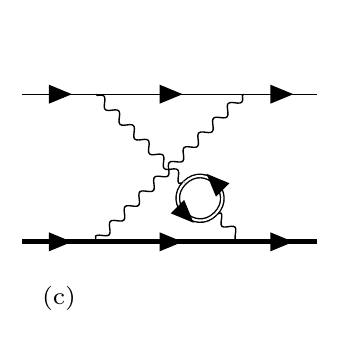}\hspace{0.1cm}
          \includegraphics[width=0.23\textwidth]{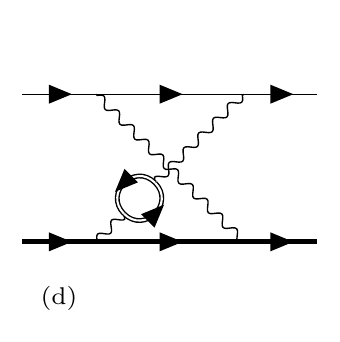}
        \caption{
          NNLO Box diagrams. Direct diagrams (a)-(b) and 
          crossed diagrams (c)-(d).}
\label{Fig:irred-box}
\end{center}
\end{figure}
\begin{figure}[t]
      \begin{center}
        \includegraphics[width=0.23\textwidth]{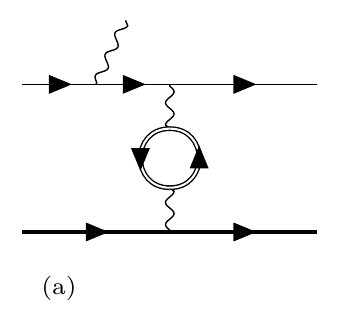}\hspace{0.1cm}
        \includegraphics[width=0.23\textwidth]{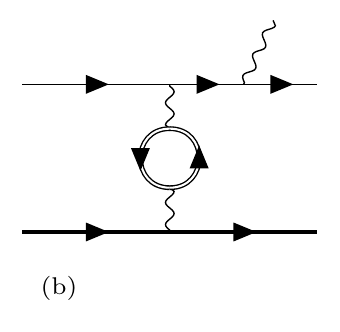}\hspace{0.1cm}
        \includegraphics[width=0.23\textwidth]{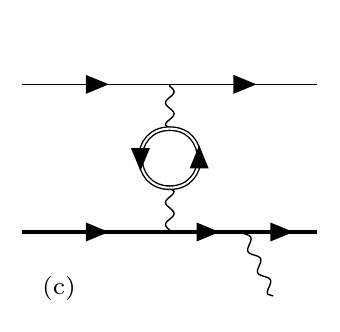}\hspace{0.1cm}
        \includegraphics[width=0.23\textwidth]{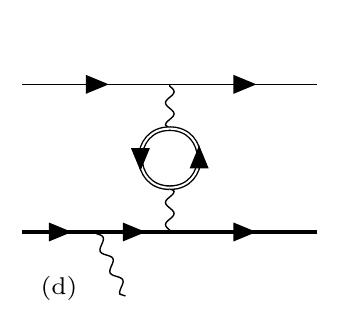}
      \caption{
        NNLO real photon radiation diagrams.
        Radiation from the electron line (a)-(b) and 
        from the muon line (c)-(d).}
        \label{Fig:real-self}
\end{center}
    \end{figure}
      \begin{figure}[ht]
        \begin{center}
          \includegraphics[width=0.23\textwidth]{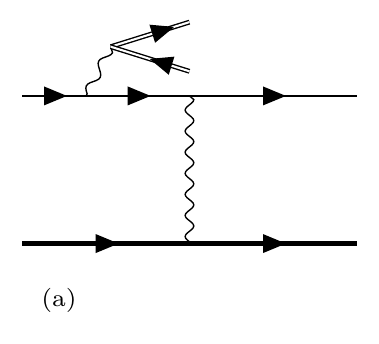}\hspace{0.1cm}
          \includegraphics[width=0.23\textwidth]{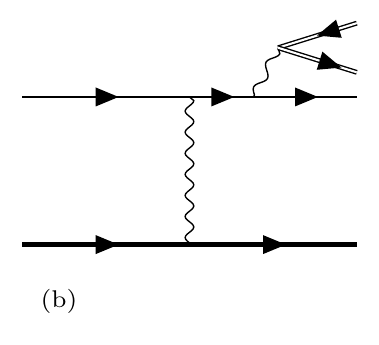}\hspace{0.1cm}
          \includegraphics[width=0.23\textwidth]{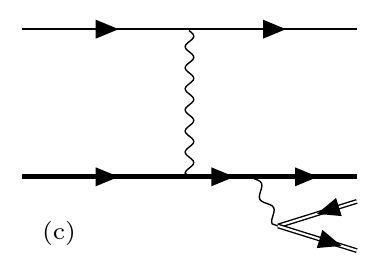}\hspace{0.1cm}
          \includegraphics[width=0.23\textwidth]{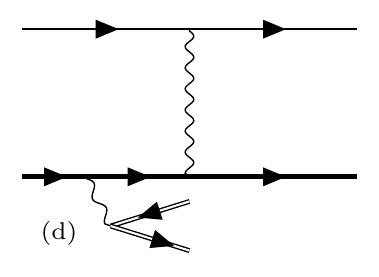}\\
          \includegraphics[width=0.23\textwidth]{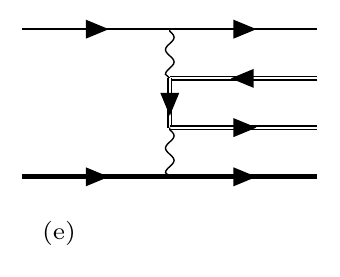}\hspace{0.5cm}
          \includegraphics[width=0.23\textwidth]{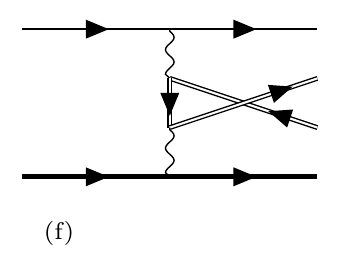}
        \caption{
        Lepton-pair real radiation diagrams.
        We do not display the diagrams obtained by exchanging the
        fermionic lines corresponding to the same lepton.
        }
        \label{Fig:real-from-electron}
        \end{center}
\end{figure}

We detail below the content of the three contributions of eq.~(\ref{eq:class-def}):
\begin{itemize}
  \item  $d\sigma_\text{virt}^{\alpha^2}$, the virtual two-loop contribution,
  consists of several classes, both factorisable and non-factorisable:
  \begin{enumerate}
    \item \label{itm:LOm2} the squared absolute value of the
      leading-order (LO) amplitude ${\mathcal A}_0$ with one-loop
      photon vacuum polarisation (VP) due to leptons
      (figure~\ref{Fig:lepton-one-loop} (b));
    \item \label{itm:2vp} interference between LO diagram and two-loop diagrams
      with iterated photon VP $\Delta\alpha^\text{LO}$ as well as two-loop irreducible
      VP $\Delta\alpha^\text{NLO}$ (figure~\ref{Fig:vacpol2}).
 
The sum of the two contributions in items~\ref{itm:LOm2}
and~\ref{itm:2vp} is factorisable, {\it i.e.} it is equal to
$\left[3\left(\Delta\alpha^\text{LO}(t)\right)^2 +
  2\Delta\alpha^\text{NLO}(t)\right]|{\mathcal A}_0|^2$, where $t$ is
the space-like four momentum of the virtual photon flowing in
figure~\ref{Fig:lepton-one-loop} (a);
\item \label{itm:LOvpx1L} interference between the one-loop diagram of
  figure~\ref{Fig:lepton-one-loop} (b) and one-loop virtual photonic
  corrections (figure~\ref{Fig:one-loop-phot}). If the amplitude for
  the latter is indicated as ${\mathcal A}_\text{1L}$, this
  contribution is factorisable and simply equal to
  $2\Delta\alpha^\text{LO}(t)\;\operatorname{Re}({\mathcal
    A}_0^\dagger{\mathcal A}_\text{1L})$;

\item \label{itm:vp_vertex} interference between LO diagram and
  two-loop factorisable contributions where the one-loop diagrams with
  photonic corrections include a VP insertion in the non-loop photon
  propagator. Only the two diagrams of figure~\ref{Fig:red-vert}
  belong to this class, which exhibits IR divergences. Also
  for this case the function $\Delta\alpha^\text{LO}(t)$ can be
  factored out;

\item interference between LO diagram and two-loop irreducible
  contributions where the one-loop diagrams with photonic corrections
  include a VP insertion in the loop photon propagator. We have two
  classes of diagrams:
      \begin{enumerate} 
      \item[a)] vertex corrections, which are IR safe
        (figure~\ref{Fig:irred-vert}). These diagrams contribute to
        the NNLO corrections to the QED vertex form factor;

      \item[b)] box corrections, which are IR divergent
        (figure~\ref{Fig:irred-box}). The sum of these corrections,
        the ones in items~\ref{itm:LOvpx1L} and~\ref{itm:vp_vertex}
        cancel the IR divergences of the sub-set
        $d\sigma^{\alpha^2}_\gamma$, described below;
      \end{enumerate}
  \end{enumerate}
\item $d\sigma_\gamma^{\alpha^2}$ includes the interplay between real
  photon radiation and leptonic loop insertions, in particular the
  interference between tree-level diagrams for the process $\mu^\pm
  e^- \to \mu^\pm e^- \gamma$ and the same class of diagrams with
  insertion of the leptonic loop on the virtual photon
  (figure~\ref{Fig:real-self}). IR divergences stemming from this set
  of corrections are cancelled by a sub-set of the contributions
  considered in the previous item;
    
  \item $d\sigma_\text{real}^{\alpha^2}$ contains the ${\cal
    O}(\alpha^2)$ tree-level amplitudes for the processes $\mu^\pm e^-
    \to \mu^\pm e^-\ell^+\ell^-$, with $\ell= e, \mu$. The production
    of the $\tau^+ \tau^-$ pair is forbidden by the available centre
    of mass energy in the MUonE experiment.  We have three different
    gauge-invariant classes:
    \begin{enumerate}
    \item lepton-pair production through photon radiation from the
      electron line (figure~\ref{Fig:real-from-electron} (a) and (b)),
    \item lepton-pair production through photon radiation from the
      muon line (figure~\ref{Fig:real-from-electron} (c) and (d)) and
    \item peripheral diagrams, depicted in
      figure~\ref{Fig:real-from-electron} (e) and (f), where the
      additional leptonic pair originates from the virtual photon
      propagator~\footnote{The plots displayed in
      figures~\ref{Fig:real-from-electron} do not include the diagrams
      obtained through exchange of identical electrons or muons, which
      are, however, included in the calculation.}.
    \end{enumerate}
\end{itemize}    

The real-pair radiation contribution $d\sigma_\text{real}^{\alpha^2}$
is particularly important for two reasons: on the pure
phenomenological side, when two final state particles escape the MUonE
acceptance, the 4-fermion final state mimics a signal event and
therefore its contribution is a reducible background and has to be
thoroughly investigated under typical MUonE running conditions. On the
other side, the interference between different real pair radiation
diagrams is expected to give a partial cancellation with its virtual
counterpart. For instance, it is known from the literature on Bhabha
scattering~\cite{Jadach:1992nk,Jadach:1993wk,Arbuzov:1995cn,Arbuzov:1995vj,Jadach:1996ca,Montagna:1998vb,Montagna:1999eu,CarloniCalame:2011zq}
that the interference between diagrams (a) and (b) of
figure~\ref{Fig:real-from-electron}, once integrated over the phase
space available to the electronic pair, develops a term proportional
to $\alpha^2\log^3(-t/m^2)$, which cancels when added to the virtual
contribution originating from diagram (a) of
figure~\ref{Fig:irred-vert}.  Similar partial cancellations are
expected between virtual contributions and the interference between
the diagrams corresponding to the cut of the closed leptonic loop.
Moreover, a careful study of the impact of the peripheral diagrams is
necessary in view of their potentially large contribution, as will be
discussed in the following sections.

\section{Calculation of virtual and real-virtual corrections}
We build our NNLO virtual leptonic corrections by means of dispersion
relation (DR) techniques, starting from the available NLO calculation
implemented in~\textsc{Mesmer}. According to
refs.~\cite{Hoang:1995ex,Actis:2007fs,Actis:2008sk,Kuhn:2008zs}, and
the seminal work of refs.~\cite{Barbieri:1972as,Barbieri:1972hn}, an
amplitude involving a photon in a loop with a VP insertion, due to
lepton $\ell$, is obtained by replacing the photon propagator as
follows
\begin{equation}
  \frac{-i g_{\mu \nu}}{q^2 + i \epsilon} \to
  \frac{-i g_{\mu \delta}}{q^2 + i \epsilon}
  i\left(q^2 g^{\delta \lambda} - q^\delta q^\lambda \right)
  \Pi_\ell(q^2)\frac{-i g_{\lambda \nu}}{q^2 + i \epsilon}\, ,
  \label{Eq:photon-prop-subst}
\end{equation}
where the renormalised VP function $\Pi(q^2)$ is obtained
from its imaginary part by means of the subtracted DR
\begin{equation}
  \Pi_\ell(q^2)= -\frac{q^2}{\pi}\int_{4 m_\ell^2}^{\infty} \frac{dz}{z}
  \frac{\operatorname{Im}\;\Pi_\ell(z)}{q^2 - z + i \epsilon}\,
  \label{Eq:subtrDR}
\end{equation}
with
\begin{eqnarray}
  \text{Im}\Pi_\ell(z) &=& - \frac{\alpha}{3} R_\ell(z) \, , \\
  R_\ell(z) &=& \left( 1 + \frac{4 m_\ell^2}{2 z}\right)
  \sqrt{1 - \frac{4 m_\ell^2}{z}} 
  \label{{Eq:leptonic-Pi}}\, ,
\end{eqnarray}
$m_\ell$ representing the mass of the lepton circulating in the loop.
Since the $q^\delta q^\lambda$ term in eq.~(\ref{Eq:photon-prop-subst})
does not contribute because of gauge invariance, the two-loop
calculation can be performed starting from the one-loop
amplitude with the replacement
\begin{equation}
  \frac{-i g_{\mu \nu}}{q^2 + i \epsilon} \to - i g_{\mu \nu} \left(
  \frac{\alpha}{3 \pi} \right) \int_{4 m_\ell^2}^\infty \frac{dz}{z}
  \frac{1}{q^2 - z + i \epsilon}\left( 1 + \frac{4 m_\ell^2}{2 z}\right)
  \sqrt{1 - \frac{4 m_\ell^2}{z}}\, .
  \label{Eq:leptonic-subtrDR}
\end{equation}
The above approach has been used in
ref.~\cite{Actis:2007fs,Actis:2008sk,Kuhn:2008zs,CarloniCalame:2011zq}
for the calculation of two-loop QED hadronic and leptonic corrections
to Bhabha scattering.  More recently the same technique has been used
in refs.~\cite{Fael:2018dmz,Fael:2019nsf} to calculate the NNLO
hadronic corrections to $\mu e$ scattering in the context of the MUonE
experiment. Partial preliminary results on NNLO virtual leptonic
corrections to muon-electron scattering have been obtained in
ref.~\cite{Balzani:2020}.

Equation~(\ref{Eq:leptonic-subtrDR}) leads in a natural way to perform
a MC integration in $dz$, for the evaluation of NNLO virtual
pair corrections, starting from the NLO calculation. Indeed, by
exchanging the order of integration variables between the loop
momentum and the effective photon mass $\sqrt{z}$ given by the
DR, the one-loop amplitude ${\mathcal
  A}_\text{1L}(\mu^\pm e^- \to \mu^\pm e^-)$ can be convoluted with a
kernel function to obtain the two-loop amplitude with the VP insertion
in the following way
\begin{equation}
  {\mathcal A}_\text{2L}(\mu^\pm e^- \to \mu^\pm e^-) = \left( 
  \frac{\alpha}{3 \pi} \right) \int_{4 m_\ell^2}^\infty \frac{dz}{z}
  R_\ell(z) {\mathcal A}_\text{1L}(\mu^\pm e^- \to \mu^\pm e^-; z)\, ,
  \label{Eq:DRconvolution}
\end{equation}
where ${\cal A}_\text{1L}(\mu^\pm e^- \to \mu^\pm e^-; z)$ represents
the one-loop amplitude with the mass of the virtual photon, where the
VP is inserted, set to $\sqrt{z}$.

In this way the CPU load for the two-loop calculation remains of the
same order of the one-loop calculation, having only one additional
dimension in the MC integration.  An important point to be checked is
the numerical stability of the algorithm. For this reason we performed
several internal checks on the convergence of the MC error estimate
and cross checks with available analytic expressions, which we
document below.

\section{Calculation of real corrections}
\label{sec:real-pair-calculation}
The calculation of the differential and integrated cross sections for
real pair emission, i.e. for the processes $\mu^\pm e^- \to \mu^\pm
e^- \ell^+ \ell^-$, with $\ell = e, \mu$ (the case $\ell = \tau$ is
not allowed at MUonE by energy conservation) is a standard tree-level
calculation for $2 \to 4$ processes.  Because of the presence of the
small electron mass scale and the need of covering the full available
phase space in the simulation, the matrix elements have to be computed
in a numerically stable way and the numerical integration has to be
performed with dedicated samplings, in particular for the final state
with electronic pair.  As will be discussed later, also the scattering
amplitudes for the processes $\mu^\pm e^- \to \mu^\pm e^- \tau^+
\tau^-$ with $m_\tau = m_e$ has been calculated.

The matrix elements (considering only QED interactions) have been
calculated with the help of the symbolic manipulation program
\textsc{Form}~\cite{Vermaseren:2000nd,Kuipers:2012rf,Ruijl:2017dtg}
and cross-checked with the automatic package
\textsc{Recola}~\cite{Actis:2016mpe,Denner:2017wsf}, finding perfect
agreement (typically the numerical relative difference is at the $10^{-12}$ level and never worse than $10^{-8}$)~\footnote{In order to select subsets of
gauge invariant classes of diagrams, for the calculation of the
amplitude $\mu^\pm e^- \to \mu^\pm e^- \tau^+ \tau^-$ with $m_\tau =
m_e,m_\mu$ with \textsc{Recola}, we used the rescaling factors
$\lambda_\mu$ and $\lambda_\tau$ for the $\mu \mu \gamma$ and $\tau
\tau \gamma$ couplings, respectively.}.

The phase space integration has been carried out with two independent
implementations. For both of them the final state momenta are
generated in the centre of mass frame and boosted in the laboratory
frame. According to one implementation, the 4-body Lorentz invariant
phase space
\begin{equation}
dLips = \int \frac{d^3 p_3}{(2 \pi)^3 2 E_3} \frac{d^3 p_4}{(2 \pi)^3 2 E_4} 
\frac{d^3 p_5}{(2 \pi)^3 2 E_5} \frac{d^3 p_6}{(2 \pi)^3 2 E_6} \delta^4 \left( p_1+p_2 - \sum_{i=3}^6 p_i\right)
\end{equation}
is decomposed into different parameterisations.  The diagrams (a) and
(b) of figure~\ref{Fig:real-from-electron}, with the lepton pair
radiated from the electron line, require the following
parameterisation:
\begin{eqnarray}
dLips &=& (2 \pi)^3 \int dQ^2 d\Phi_3(P \to p_3 + p_4 + Q)\,  d\Phi_2(Q \to p_5 + p_6)  
\end{eqnarray}
where $P = p_1+p_2$ is the total initial state momentum (with $p_1$
and $p_2$ being the initial muon and electron momenta, respectively),
$d\Phi_n$ is the $n$-body phase space element
\begin{equation}
d\Phi_n = \int \prod_{i=1}^n\frac{d^3 p_i}{(2 \pi)^3 2 E_i} \delta^4 \left( P - \sum_{i=1}^n p_i\right)\, 
\end{equation}
and $p_3$, $p_4$, $p_5$ and $p_6$ are the final state muon, electron
and $\ell^+$, $\ell^-$ momenta, respectively.  In order to obtain
reliable numerical MC predictions, different sets of independent
variable mappings are necessary to tailor the generation according to
the peaking behaviour of diagrams (a) and (b) of
figure~\ref{Fig:real-from-electron}.  In particular, for diagram (a)
the following set of independent variables is used:
\begin{equation}
Q^2, \cos\theta_4, \phi_4, \cos\theta_{56},  \phi_{5}, E_{56}, \cos\theta_5^*, \phi_5^*\,,
\end{equation}
where $\theta_3^*$ and $\phi_3^*$ are generated in the rest-frame of
the pair ${56}$ $(\vec{q_5} + \vec{q_6} = \vec{0})$, and $\theta_{56}$
and $\phi_{56}$ are the polar and azimuthal angles of the pair
momentum $(q_5+q_6)$ in the centre of mass frame.  For diagram (b),
$\theta_{56}$ and $\phi_{56}$ are understood in the centre of mass
frame with $z$-axis oriented along the outgoing electron $\vec{p}_4$.
The diagrams (c) and (d) of figure~\ref{Fig:real-from-electron}, with
the lepton pair radiated from the muon line, do not require dedicated
parameterisation because $m_\mu$ is of the same order of magnitude of
$\sqrt{s}$ and therefore the logarithmic structure developed by the
numerical integration is not leading.  The peripheral diagrams (e) and
(f) of figure~\ref{Fig:real-from-electron} require a different phase
space parameterisation~\cite{Berends:1994pv,Kersevan:2004yh}:
\begin{eqnarray}
dLips &=& (2 \pi)^6 \int dQ_{356}^2\,  dQ_{56}^2 \nonumber \\ 
&\times& d\Phi_2(P \to p_4 + Q_{356})\,  d\Phi_2(Q_{356} \to p_3 + Q_{56})\, d\Phi_2(Q_{56}\to p_5 + p_{6})\, .
\end{eqnarray}
The independent variables are the invariant 
masses 
\begin{equation}
Q_{356}^2, Q_{56}^2
\end{equation}
and the angles 
\begin{equation}
\cos\theta_4\, , \phi_4\, , \cos\theta_3^*\, , \phi_{3}^*\, , \cos\theta_5^{**}\, , \phi_5^{**}\,,
\end{equation}
where $\cos\theta_3^*$, $\phi_{3}^*$ 
are generated in the $Q_{356}$ rest-frame and 
$\cos\theta_5^{**}$ and $\phi_{5}^{**}$ are generated in the $Q_{56}$ rest-frame.

When the $\ell^+ \ell^-$ pair is an electron-positron pair $e^+ e^-$,
all the necessary simmetrised channels with $p_4$ and $p_6$ exchanged
are considered.

A second (and independent) phase-space decomposition has been
implemented, which is the one used in the official release of
\textsc{Mesmer}. In this case, in the principal integration channel,
the independent variables are
\begin{equation}
  Q_{456}^2,\, Q_{56}^2, t_{13},\,\phi_3,\, \cos\theta_{56}^\dagger,\,
  \phi_{56}^\dagger,\, \cos\theta_5^\star,\, \phi_5^\star,
\label{eq:myindipvars}
\end{equation}
where $Q_{456}^2 = (p_4+p_5+p_6)^2$, $Q_{56}^2 = (p_5+p_6)^2$,
$t_{13}=(p_1-p_3)^2$, $\phi_3$ is the $p_3$ azimuthal angle in the
$p_1+p_2$ rest frame, $\theta_{56}^\dagger$ and $\phi_{56}^\dagger$
are the $p_5+p_6$ angles in the $p_1-p_3 +p_2$ rest frame, and
$\theta_{5}^\star$ and $\phi_{5}^\star$ are the $p_5$ angles in the
$p_5+p_6$ rest frame. The importance sampling of the variables in
eq.~\ref{eq:myindipvars} follows as close as possible the peaking
structure of the differential cross section, and multi-channel
techniques are used to map different structures.
For instance, in the case of an extra electronic pair and referring to
figure~\ref{Fig:real-from-electron}, in the
principal channel, while $\phi_3$, $\phi_{56}^\dagger$ and
$\phi_5^\star$ are sampled uniformly, $Q_{456}^2$ is sampled as
$1/Q_{456}^2$ to follow the singularity due to the internal fermion
propagator of diagram (b),
$t_{13}$ is sampled according to $1/t_{13}$ to flatten the peaks of
the photon propagator exchanged between the muon and electron lines and
$Q_{56}^2$ is sampled following $1/Q_{56}^2$. The cosine
$\cos\theta_{56}^\dagger$ is distributed according to
$1/(1-\beta_{2}\cos\theta_{56}^\dagger)$ or 
$1/(1-\beta_{13}\cos\theta_{56}^\dagger)$, where
$\beta_{2}$ ($\beta_{13}$) is the speed of $p_2$ ($p_1-p_3$) in
the $p_1-p_3 +p_2$ rest frame, to flatten the peaks due to the
internal electron propagator of diagram (a)
or the photon propagator in
diagram (e). Finally, $\cos\theta_5^\star$ is sampled according to
$1/(1-\beta_{13}\cos\theta_5^\star)$ or $1/(1-\beta_{24}\cos\theta_5^\star)$, where 
$\beta_{13}$ ($\beta_{24}$) is the speed of $p_1-p_3$ ($p_2-p_4$)  in
the $p_5+p_6$ rest frame, to flatten the peak due to the electron
propagator in diagrams (e) and (f). In the other channel,
the roles of $p_1$ and $p_3$ are interchanged with $p_2$ and
$p_4$. For identical particles, a symmetrisation of the channels is
performed by exchanging $p_4\leftrightarrow p_6 $.

\section{Numerical results}
\label{sec:numerics}
We are going to study the numerical impact of QED NNLO leptonic pair
corrections to observables for virtual and real corrections
separately, in order to understand the features and the size of each
contribution.

The simulations are performed assuming a $150\text{
  GeV}$ muon beam~\footnote{Lately, it became more likely that MUonE running
conditions will use a $160\text{ GeV}$ muon beam. Here we still use a
$150\text{ GeV}$ energy for direct comparison with past
phenomenological results and because it does not change our
conclusions.} and according to the following event selection
criteria~\footnote{It corresponds to Setup~1 in
ref.~\cite{CarloniCalame:2020yoz} and to Setup~2 in
ref.~\cite{Alacevich:2018vez}. In the latter, also setups with
$E_e>0.2$~GeV were considered, which are disregarded here for the sake
of brevity.}:
\begin{itemize}
\item $\theta_e,\theta_\mu<100\text{\ mrad}$ and $E_e > 1$~GeV.
\end{itemize}
The angular cuts model the typical acceptance conditions of the
experiment and the electron energy threshold is chosen in order to
single out a region where the signal of the experiment is not
negligible.

We remark that this event selection is driven by the $2\to 2$
kinematics and can be rephrased differently to describe the events
more in detail. In the $2\to2$ kinematics, the outgoing muon is
characterised by an angle $\theta_\mu<\bar{\theta}_\mu =
\arccos\left(\sqrt{1-m_e^2/m_\mu^2}\right)\simeq 4.84\text{ mrad}$
and, for a $150\text{ GeV}$ incident muon, an energy
$E_\mu>\bar{E}_\mu\simeq 10.28\text{ GeV}$~\footnote{It corresponds to
the laboratory energy of a muon scattered backward in the
centre-of-mass frame.}.  When we consider real lepton pair radiation
we can have more charged tracks in the detector~\footnote{In its
present proposal, MUonE will not distinguish positive from negative
charges.} and therefore we need to introduce a more general event
selection inspired by the one of the elastic $2 \to 2$ process. In
particular, we can define a muon-like track, characterised by angle
lower than $\bar{\theta}_\mu$ and energy larger than $\bar{E}_\mu$,
and an electron-like track, with angle lower than $100\text{ mrad}$
and energy larger than $1\text{ GeV}$.  The generalised selection
criteria can be defined as follows:
\begin{itemize}
  \item two tracks in the detector, one muon-like
    ($\theta_\mu<\bar{\theta}_\mu\simeq 4.84\text{ mrad}$ and
    $E_\mu>\bar{E}_\mu\simeq 10.28\text{ GeV}$) and one
    electron-like ($\theta_e<100\text{ mrad}$ and $E_e>1\text{ GeV}$)
\end{itemize}
This alternative definition will be used in the following when 4-body
final states are studied, and we will name it {\it basic acceptance
  cuts}.

On top of it, we are going to consider also three extra selection cuts
(and combinations of them), which are tailored to select elastic
events in order to suppress reducible backgrounds and keep the
sensitivity to the signal. The elasticity cuts we implement are:
\begin{enumerate}
\item \texttt{cut 1} $-$ an additional cut on the minimum electron-
  and muon-like scattering angles, $\theta_{e},\theta_{\mu} >
  \theta_c$, to reduce the impact of peripheral diagrams in events
  with additional electron pair emission.  For illustrative purposes,
  we use the value $\theta_c = 0.2\text{ mrad}$, which appears to be a
  good compromise between the requirement of keeping sensitivity to
  $\Delta\alpha(t)$ in the large $\vert t \vert$ region~\footnote{We
  remind the reader that the largest sensitivity occurs for $\theta_e
  < 5$~mrad (see figure~13 of ref.~\cite{Alacevich:2018vez}).} and
  effectively suppressing the reducible background;

\item \texttt{cut 2} $-$ acoplanarity cut $\xi =
  \left|\pi-\left|\phi_e-\phi_\mu\right|\right|< \xi_c = 3.5 \,
  \text{mrad}$~\footnote{The value $\xi_c = 3.5 \, \text{mrad}$ is
  chosen in analogy with
  refs.~\cite{Alacevich:2018vez,CarloniCalame:2020yoz} and perhaps it
  is too small for MUonE running conditions.}, where $\phi_{e,\mu}$
  are the azimuthal angles of the electron- and muon-like momenta;

\item \texttt{cut 3} $-$ elasticity distance $\delta < \delta_c =
  0.2\text{ mrad}$, where $\delta$ is defined as the minimal Cartesian
  distance of the $(\theta_e$, $\theta_\mu)$ point in the
  $\theta_e$-$\theta_\mu$ plane from the elastic curve. The elastic
  curve can be parameterised by the function
    \begin{equation}
    \theta_\mu(\theta_e) = \arctan{\left[\frac{2 m_e r \cos{\theta_e} \sin{\theta_e}}{E_\mu^i - r\left( r E_\mu^i + 2 m_e\right) \cos^2{\theta_e}}\right]}\, ,
    \label{Eq:elastic-corr}
    \end{equation}
    where $r$ is defined in eq.~(9) of ref.~\cite{Abbiendi:2016xup}
    and $E_\mu^i$ is the incident muon energy in the laboratory
    reference frame. The distance of the generic point
    $(\theta_e^0,\theta_\mu^0)$ is defined as
    \begin{equation}
    \delta= \min_{\theta_e}\sqrt{(\theta_e - \theta_e^0)^2 + ( \theta_\mu(\theta_e)-\theta_\mu^0)^2}\, .
    \label{eq:distance}
    \end{equation}
    While the optimal value for $\delta_c$ can be determined only with
    a realistic simulation that takes into account of all the
    experimental conditions and detector effects, we use for the
    present study the reference value
    $\delta_c =0.2$~mrad.
\end{enumerate}

When considering the emission of real leptonic pairs in the following,
in order to study the effects from different gauge invariant
contributions and their interplay with the virtual corrections from a
purely theoretical point of view, neglecting the effect of identical
particles, we will begin our analysis considering the processes
$\mu^\pm e^- \to \mu^\pm e^-\tau^+ \tau^-$ for the two cases $m_\tau =
m_e$ and $m_\tau = m_\mu$. The three gauge invariant contributions are
the radiation from the electron line, from the muon line and the two
peripheral diagrams, according to the discussion of
section~\ref{sec:class} on the ingredients of
$d\sigma^{\alpha^2}_\text{real}$. In this exercise we are completely
inclusive over the additional ``$\tau$'' pair and assume the cuts are
applied to the final state electron and muon, as if they were
distinguishable from the extra leptons in the final state (asymmetric
cuts). The obtained numerical results must be considered ``academic''
since the event selection does not fully match the realistic MUonE
experimental settings.

As a second step, we consider the true processes $\mu^\pm e^- \to
\mu^\pm e^- e^+ e^-$ and $\mu^\pm e^- \to \mu^\pm e^- \mu^+ \mu^-$,
taking into account more realistic conditions.  In MUonE the final
state charged particles are detected through their tracks in the
silicon detectors, without charge identification. We therefore require
that exactly two tracks
are reconstructed in the detector
by asking that they have
a threshold energy larger than $E_\text{thr.}= 0.2$~GeV and an angle
lower than $\theta_\text{thr.}=100$~mrad. If more than two tracks
are reconstructed,
the event is rejected. Otherwise, of the two
selected tracks,
we require one to be muon-like and the other to be
electron-like, as per definitions above. We will further show the
effects of cutting through \texttt{cut 1-3}. We remark that in
principle any pair combination of final state leptons can fall within
the only-two-tracks criterium, but we verified {\it a posteriori} that
the final state muon of the $\mu^\pm e^-\to\mu^\pm e^-\ell^+\ell^-$
process is always the muon-like track.

The input parameters for the simulations are set to
\begin{equation}
  \alpha = 1/137.03599907430637 \qquad
  m_e = 0.510998928\text{ MeV} \qquad m_\mu = 105.6583715\text{ MeV}\nonumber.
  \label{eq:inputqedparams}
\end{equation}
If $p_{1(2)}$ is the incoming muon (electron) momentum and $p_{3(4)}$
is the muon(electron)-like outgoing momentum~\footnote{We notice that,
when considering the emission of an extra real electronic pair, $p_4$
can be any of the final state $e^-$ or the $e^+$ momentum, according
to which one represents the electron-like track.}, the considered
differential observables are the following ones:
\begin{equation}
  \frac{d\sigma}{d t_{ee}}\,, \, \, \, \, \, \,
  \frac{d\sigma}{d t_{\mu\mu}}\,, \, \, \, \, \, \,
  \frac{d\sigma}{d\theta_e}\,, \, \, \, \, \, \,
  \frac{d\sigma}{d\theta_\mu}
  \label{eq:observables}
\end{equation}
where $t_{ee} = (p_2 - p_4)^2$, $t_{\mu \mu} = (p_1 - p_3)^2$,
$\theta_e$ and $\theta_\mu$ are the scattering angles, in the
laboratory frame, of the outgoing electron- and muon-like tracks,
respectively.

We decide to represent the corrections as a differential $K$ factor defined by
\begin{equation}
K_\text{NNLO} = \frac{d\sigma^{\alpha^2}_{N_f}}{d\sigma_\text{LO}}\, ,
    \label{eq:Kfactor}
\end{equation}
{\it i.e.} we normalise the differential cross sections under study
w.r.t. the LO ones. The choice is dictated by consistency and for
direct comparison with results presented in
refs.~\cite{Alacevich:2018vez,CarloniCalame:2020yoz} and because the
10ppm accuracy goal is implicitly referred to LO results.

We notice that in the following we show also the hadronic effects
already considered in~\cite{Fael:2019nsf}, for completeness and
cross-check purposes.

\subsection{NNLO virtual pair corrections: the factorised diagrams}
\label{sec:numerics-factorized-on-born}
In this section we show the impact of NNLO virtual pair corrections of
figures~\ref{Fig:lepton-one-loop}
and~\ref{Fig:vacpol2}~\footnote{Despite they are calculated in the MC
code, here we do not show any effect due to $\tau$ or top quark loops
because they are small on the scale of the effects induced by $e$,
$\mu$ or hadronic loops.}, which are factorised over the tree-level
cross section and contribute to the running of the electromagnetic
coupling constant.
\begin{figure}[ht]
  \begin{center}
    \includegraphics[width=0.5\textwidth]{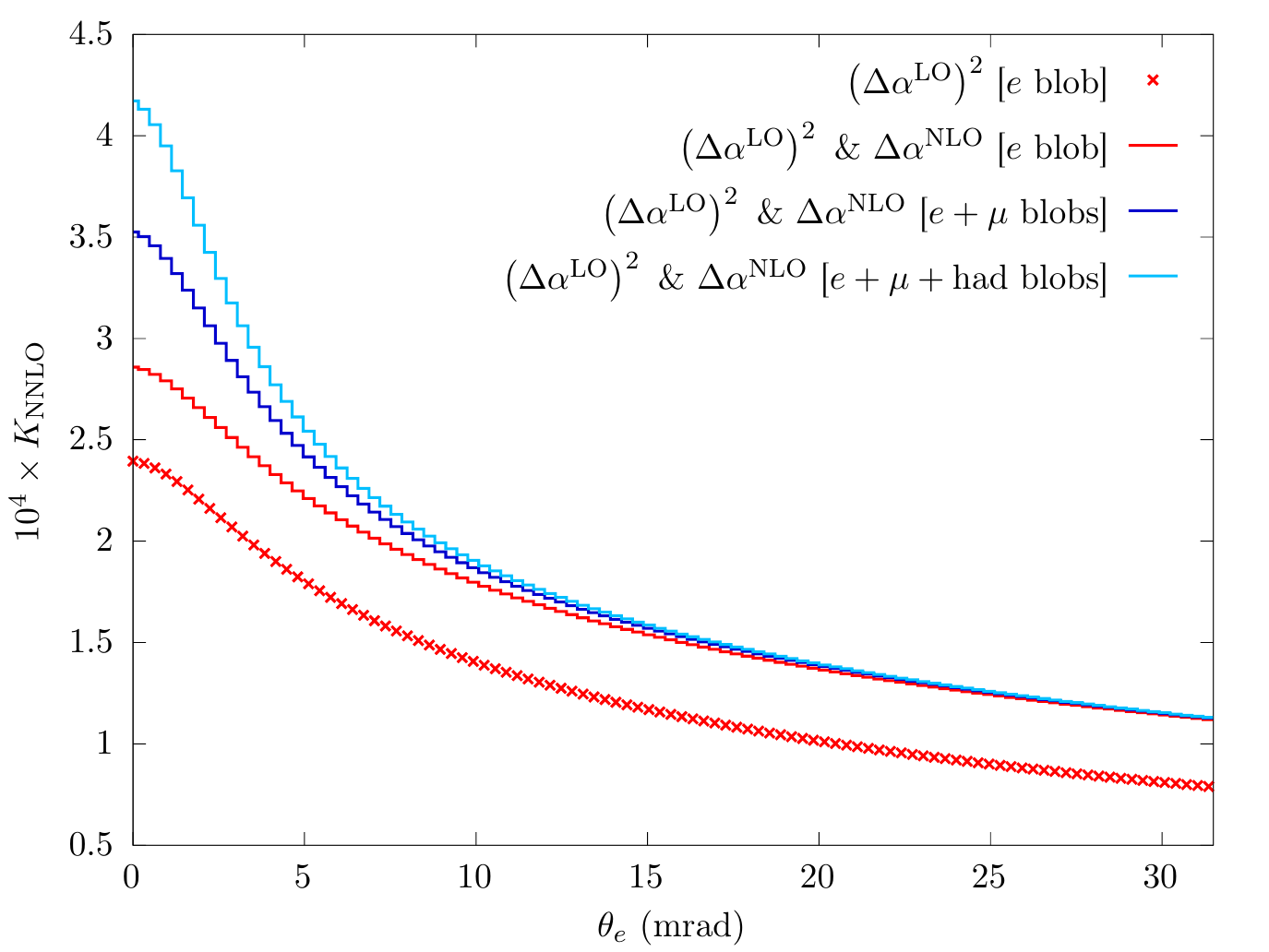}%
    \includegraphics[width=0.5\textwidth]{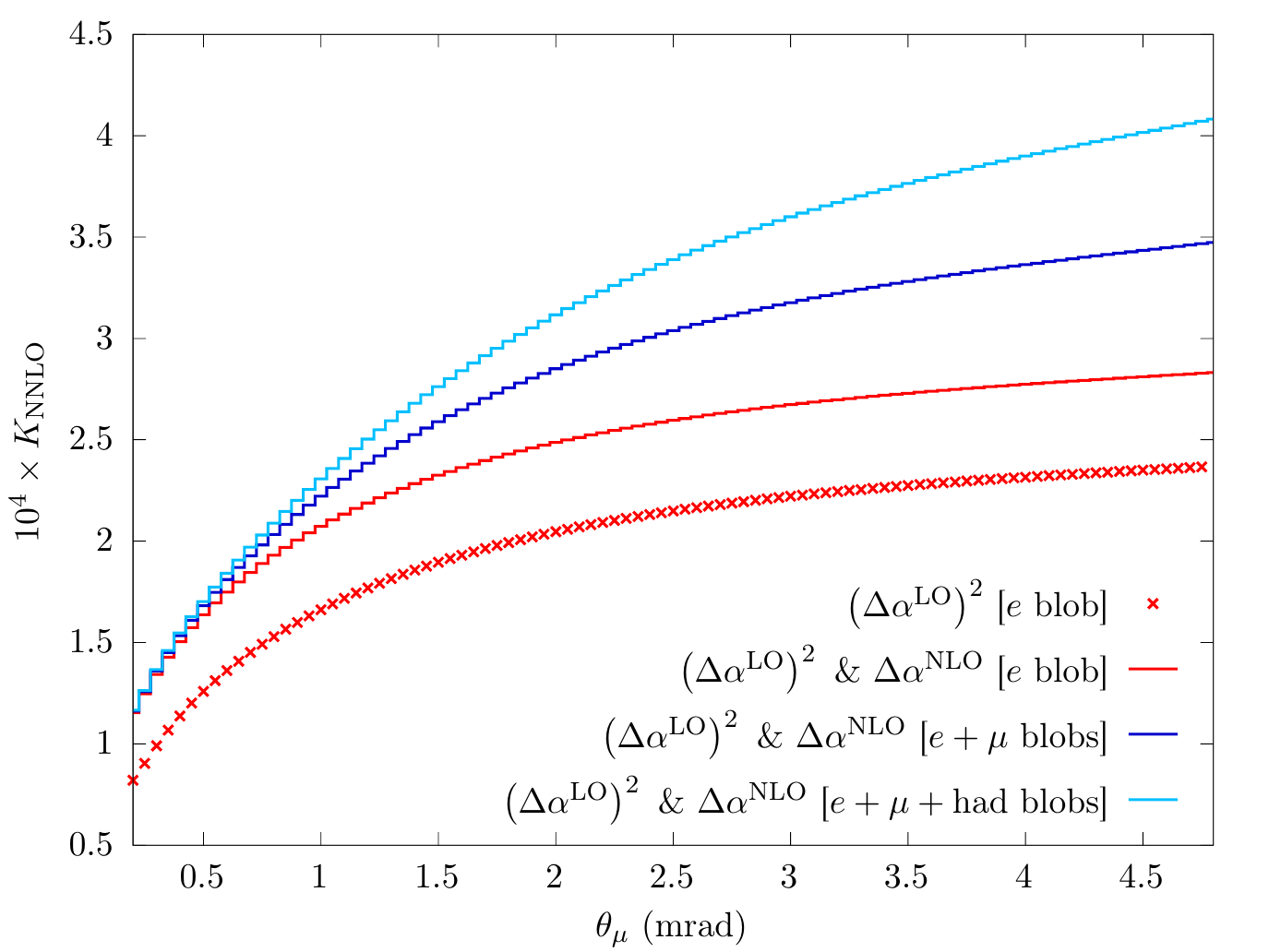}
  \caption{
    Left: differential $K_\text{NNLO}$ factor for the electron scattering angle, including the
    diagrams of figure~\ref{Fig:vacpol2}.
    Right: the same as left panel, for the muon scattering angle.
  }
  \label{Fig:vacpol2-on-thetae-thetamu}
  \end{center}
\end{figure}
In figure~\ref{Fig:vacpol2-on-thetae-thetamu}, the red histograms show
the contribution of the electron circulating in the loop on the
$\theta_e$ (left) and $\theta_\mu$ (right) distributions.  The points
stand for the term $\left(\Delta\alpha^\text{LO}\right)^2$ and solid
lines include also the term $\Delta \alpha^\text{NLO}$. The first term
is positive and ranges from about $2.4 \times 10^{-4}$ for zero
electron scattering angle to about $0.8 \times 10^{-4}$ at
$30$~mrad. As expected, the range of the corrections on $\theta_\mu$
is the same with opposite slope w.r.t. the electron angle, with the
muon angle ranging from lower to higher angles.  The irreducible
diagram of figure~\ref{Fig:vacpol2}~(b) gives a positive shift, almost
flat over the whole range of about $0.5 \times 10^{-4}$. The blue
solid line includes also the two-loop contribution with the muon
circulating in the loops.  It is a negligible contribution for
$\theta_e \gtrsim 15$~mrad ($\theta_\mu \lesssim 0.5$~mrad) and
becomes of the order of $0.6 \times 10^{-4}$ for $\theta_e \sim
0$~mrad $(\theta_\mu \to 5\,\text{mrad})$, where the exchanged $t$
increases in absolute value. The light blue solid line includes all
the leptonic contributions and the hadronic corrections, calculated
with the help of the routines of the \textsc{KNT v3.0.1}
package~\cite{Keshavarzi:2018mgv}~\footnote{Available upon request
from the authors. Detailed cross-checks with the \textsc{hadr5x19.f}
routine by Fred Jegerlehner (available at
\href{http://www-com.physik.hu-berlin.de/~fjeger/software.html}{\tt
  www-com.physik.hu-berlin.de/\~{}fjeger/software.html}, see
ref.~\cite{Jegerlehner:2018zrj}) have been carried out, finding fully
compatible results.}. The latter contribution is similar in shape and
size to the one of the muon loop.
\begin{figure}[ht]
  \begin{center}
    \includegraphics[width=0.6\textwidth]{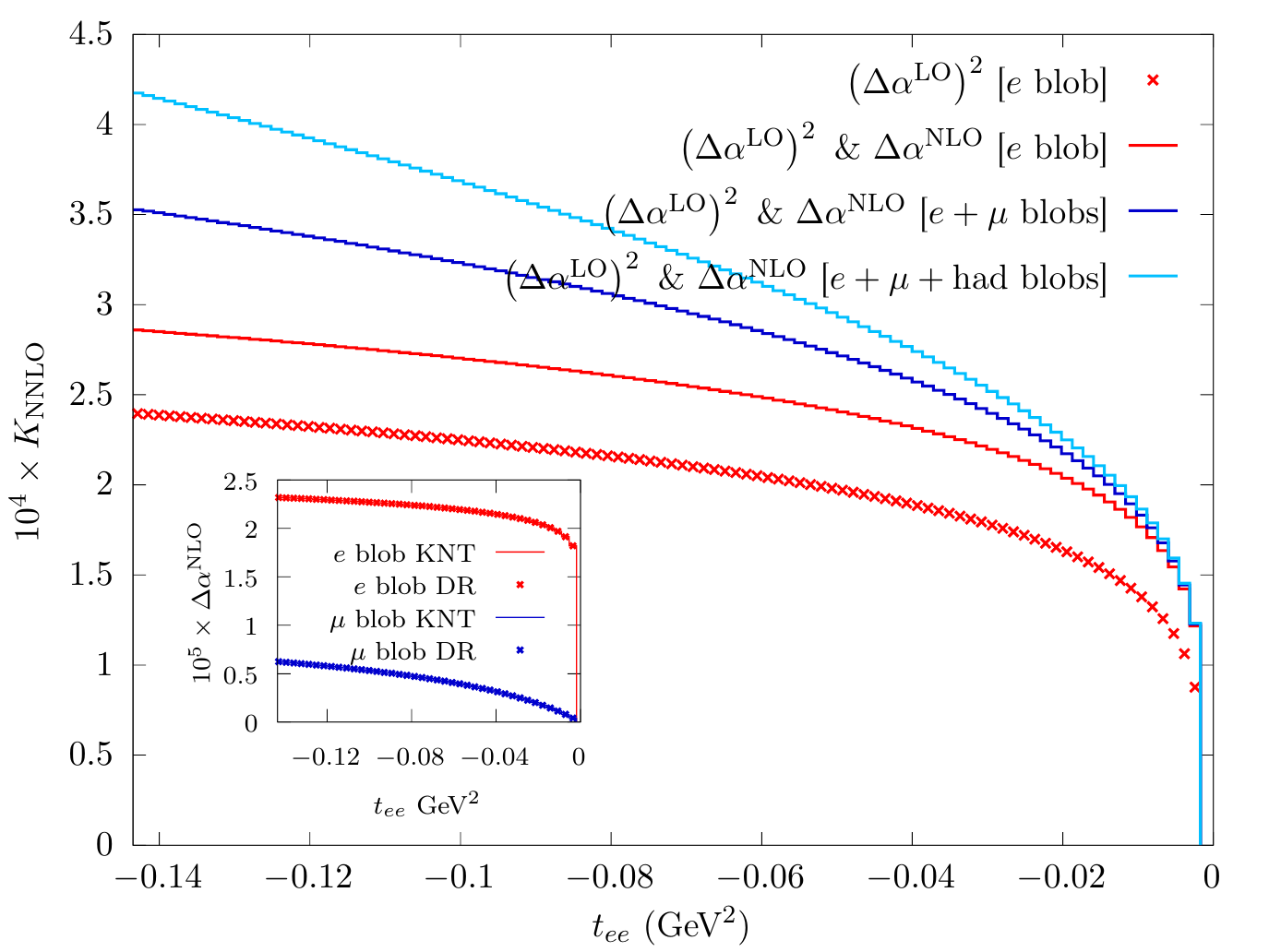}
  \end{center}
  \caption{
    Effects of the
    diagrams of figure~\ref{Fig:vacpol2} 
    on the $t_{ee}$ distribution.}
  \label{Fig:vacpol2-on-t24}
\end{figure}

The same breakdown of the separate effects, displayed in incremental
way, is shown for the $t_{ee}$ distribution in
figure~\ref{Fig:vacpol2-on-t24}.  For comparison purposes, in the
inset we present the function $10^{5}\times\Delta\alpha^\text{NLO}(t)$
for an electron or muon loop as coded in the \textsc{KNT v3.0.1}
package and as obtained through a DR, showing perfect
agreement. The DR we use here is fully analogous to
eq.~(\ref{Eq:subtrDR}) where the function $R_\ell(z)$ is replaced by
the NLO function $R^\text{NLO}_\ell(z)$ resulting from the QED limit
of eq.~(18) in ref.~\cite{Harlander:2002ur}.

\subsection{Vacuum polarisation on photonic vertex corrections}
\label{sec:numerics-vp-on-vertex}
\begin{figure}[ht]
  \begin{center}
    \includegraphics[width=0.5\textwidth]{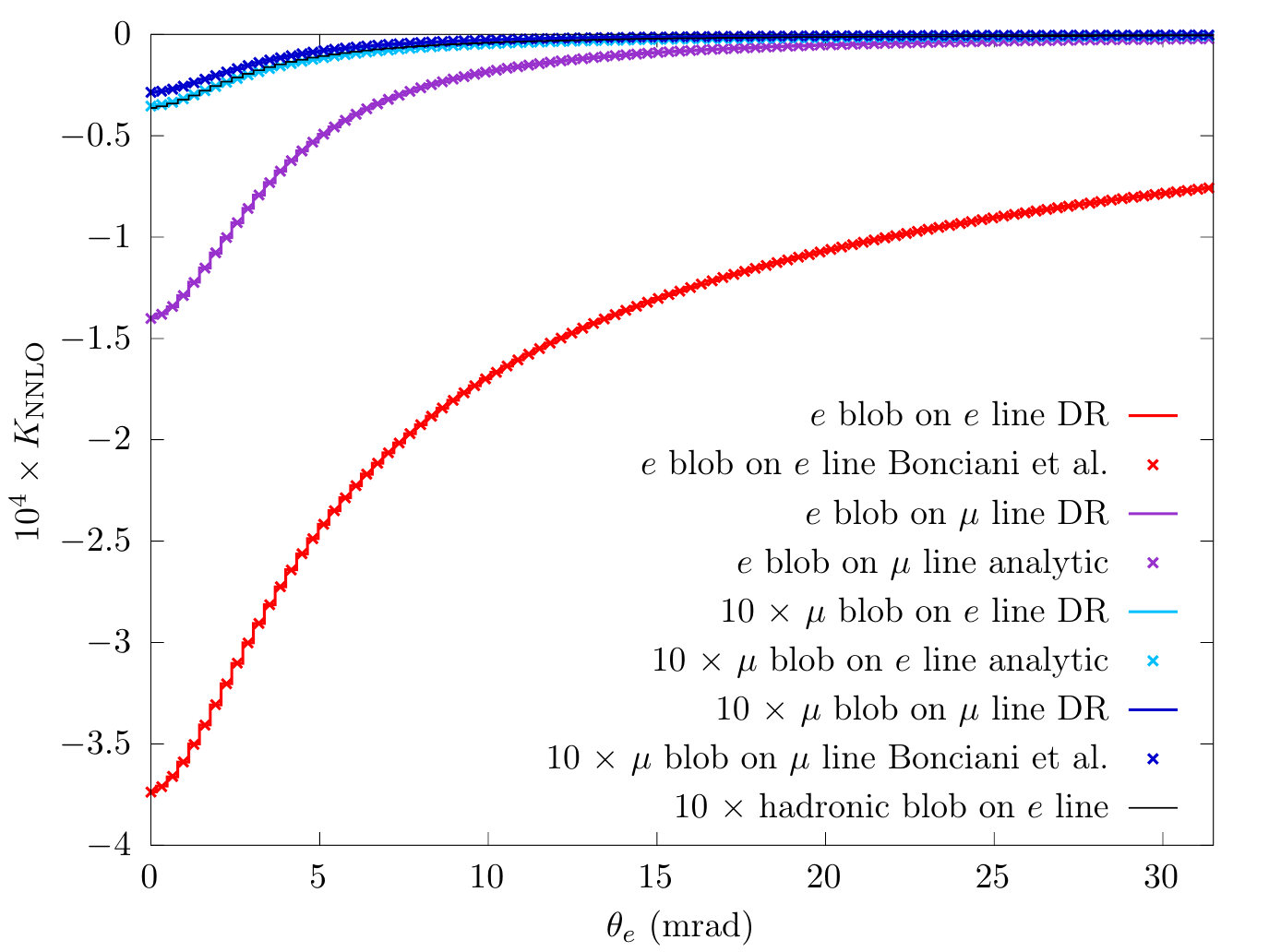}%
    \includegraphics[width=0.5\textwidth]{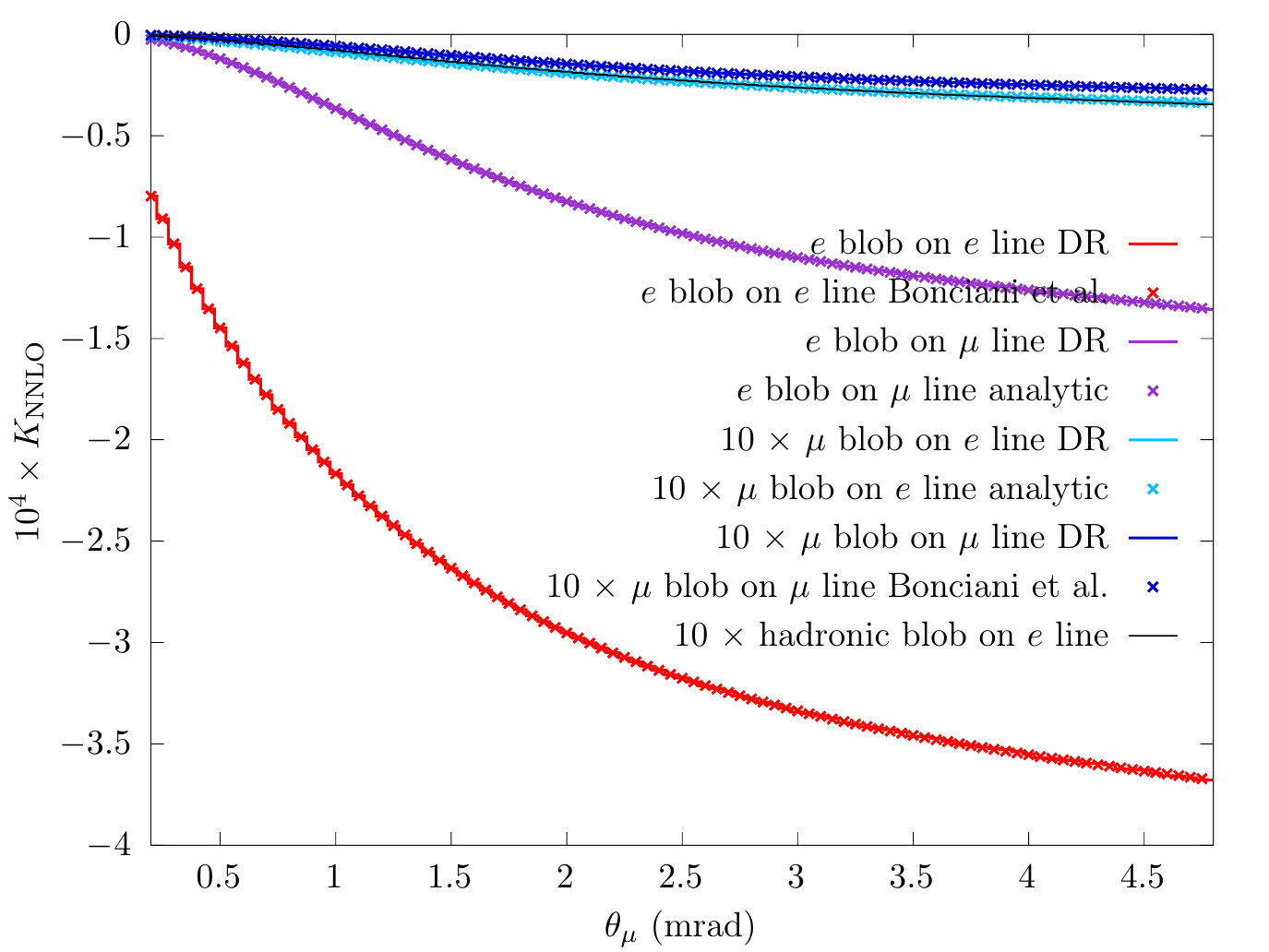}
  \caption{Left: effect of the diagrams of
    figure~\ref{Fig:irred-vert} on the $\theta_e$ distribution. The
    red histograms refer to the electron loop correction along the
    external electron line; the violet histograms show the effects of
    the electron loop correction along the external muon line; the
    light blue refer to the muon loop along the external electron
    line; the dark blue line represents the muon loop along the
    external muon line; the black line gives the hadronic loop
    contribution along the external electron line.  Right: the same as
    left panel, for the $\theta_\mu$ distribution. Notice that
    contributions due to muon or hadronic VP insertion have been
    amplified by a factor of 10 for readability on the scale of the
    plots.}
      \label{Fig:vertex-blob-on-thetae-thetamu}
  \end{center}
\end{figure}

\begin{figure}[ht]
  \begin{center}
    \includegraphics[width=0.6\textwidth]{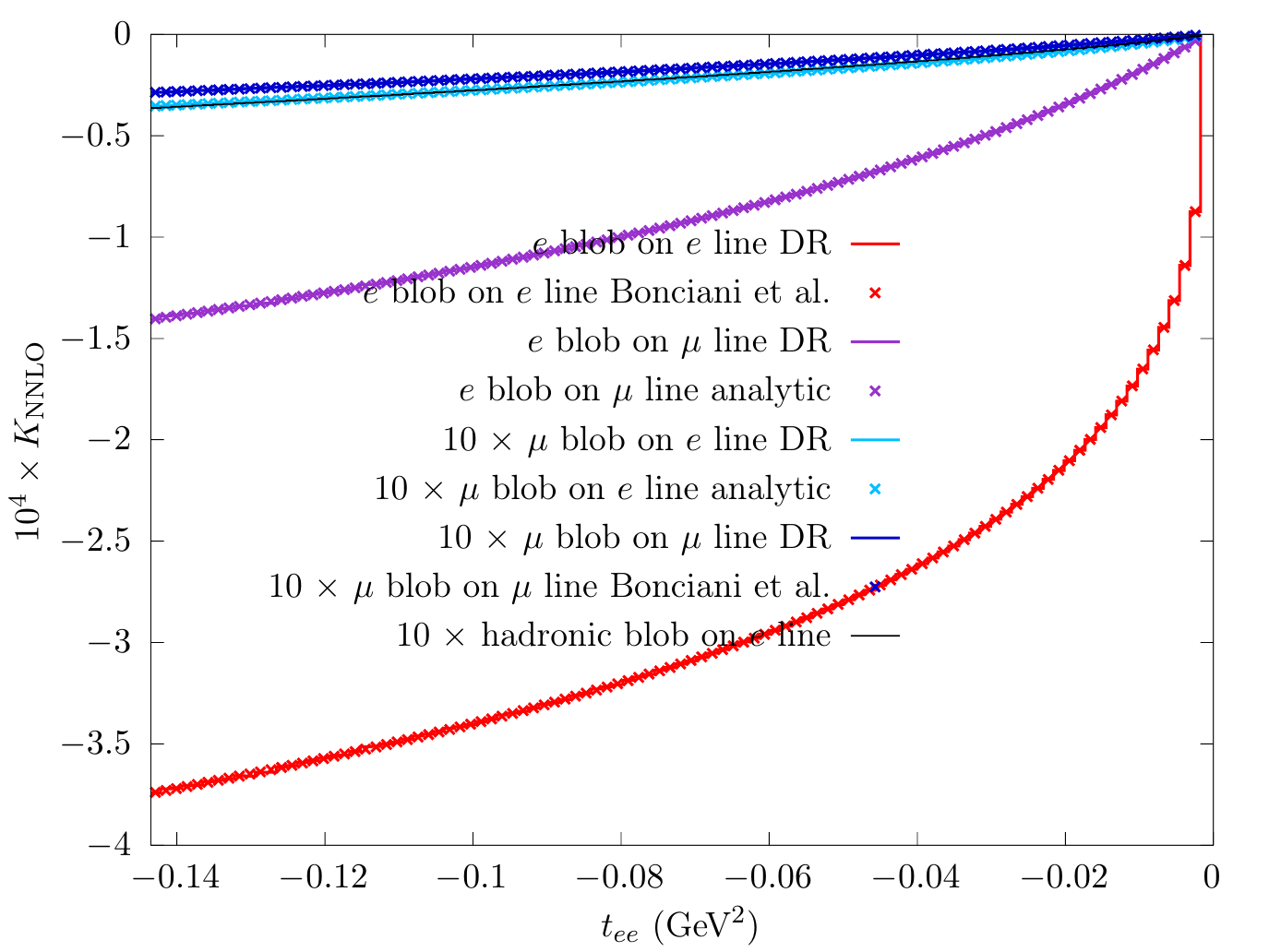}
  \caption{
  The same as figure~\ref{Fig:vertex-blob-on-thetae-thetamu} for the $t_{ee}$ distribution.
  }
  \label{Fig:vertex-blob-on-t}
  \end{center}
\end{figure}

Among the two-loop irreducible contributions where the one-loop
diagrams with photonic corrections include a VP insertion in the loop
photon propagator, the diagrams of figure~\ref{Fig:irred-vert},
together with their respective counterterms, are gauge-invariant and
IR safe.  Therefore it is meaningful to investigate their numerical
relevance.  In figures~\ref{Fig:vertex-blob-on-thetae-thetamu}
and~\ref{Fig:vertex-blob-on-t} we show, separately, all the effects of
different leptons (electron and muon) in the photonic vertex one-loop
diagrams of figure~\ref{Fig:irred-vert} for the three observables
$\theta_{e}$, $\theta_{\mu}$, $t_{ee}$.  For each contribution the
histogram refers to the calculation through DR and
the points refer to the analytical calculation. For the case of a
closed loop with a lepton with mass equal to the one of the current
which the photon loop is attached to, the analytical calculation
presented in Appendix A.2 of ref.~\cite{Bonciani:2004gi} has been
used. In all other cases ({\it i.e.} mass of the lepton flowing in the
VP insertion different from the mass of the lepton of the external
current), we use an in-house independent calculation which is
described in Appendix~\ref{sec:Syedappendix}.

We would like to remark that for all cases excellent numerical
agreement is found, which indicates also that the DR approach is implemented correctly.

From the phenomenological point of view, the largest contribution is
given, as expected, by the electron loop on the electron line (red
curves), which reaches the level of $-0.037\%$ at the kinematical
boundaries corresponding to the maximum $|t_{ee}|$. As known in the
literature, see e.g. refs.~\cite{Burgers:1985qg, Arbuzov:1995vj,
  Bonciani:2004gi}, the leading logarithmic term of the electron loop
in the vertex correction on the electron line is proportional to
$\alpha^2\log^3[{m_e^2/(-t_{ee})}]$ which is expected to be partially
cancelled by the corrections corresponding to the real emission of an
$e^+ e^-$ pair. This will be discussed in the following section with a
numerical investigation.

The electron loop on the muon line is suppressed w.r.t. the previous
contribution and reaches the level of $-0.014\%$ at the kinematical
boundary. The smallest contribution is given by the muon loop on the
muon line, which never reaches the $0.001\%$ level.


Finally, the muon loop correction on the electron line is of the same
size of the hadronic correction on the electron line, remaining at
most at the level of a few ppm.

\subsection{NNLO virtual pair corrections: the complete set of diagrams}
\label{sec:numerics-vp-on-photRC}
In this subsection we present the numerical impact of the
contributions due to the photonic one-loop amplitudes (virtual and
real contributions) with an additional closed leptonic loop, namely
the sum of the interference between the diagram in
figure~\ref{Fig:lepton-one-loop}~(a) with the diagrams in
figures~\ref{Fig:red-vert}, ~\ref{Fig:irred-vert}
and~\ref{Fig:irred-box} plus the interference of the diagram in
figure~\ref{Fig:lepton-one-loop}~(b) with the diagrams in
figure~\ref{Fig:one-loop-phot}. Also real radiation is added ({\it
  i.e.} the interference of the diagrams in figure~\ref{Fig:real-self}
with the very same diagrams without loop insertion), as well as all
the contributions described in
sections~\ref{sec:numerics-factorized-on-born}
and~\ref{sec:numerics-vp-on-vertex}. As we know from previous studies
on NLO~\cite{Alacevich:2018vez} and NNLO
photonic~\cite{CarloniCalame:2020yoz} (and also
hadronic~\cite{Fael:2019nsf}) corrections, the real radiation can give
large contributions, in particular for the electron scattering angle
distribution. For this reason in
refs.~\cite{Alacevich:2018vez,CarloniCalame:2020yoz} the additional
acoplanarity cut (\texttt{cut 2} as defined in the introduction to
this section) was introduced, in order to partially remove hard
radiation effects.
\begin{figure}[ht]
  \begin{center}
    \includegraphics[width=0.5\textwidth]{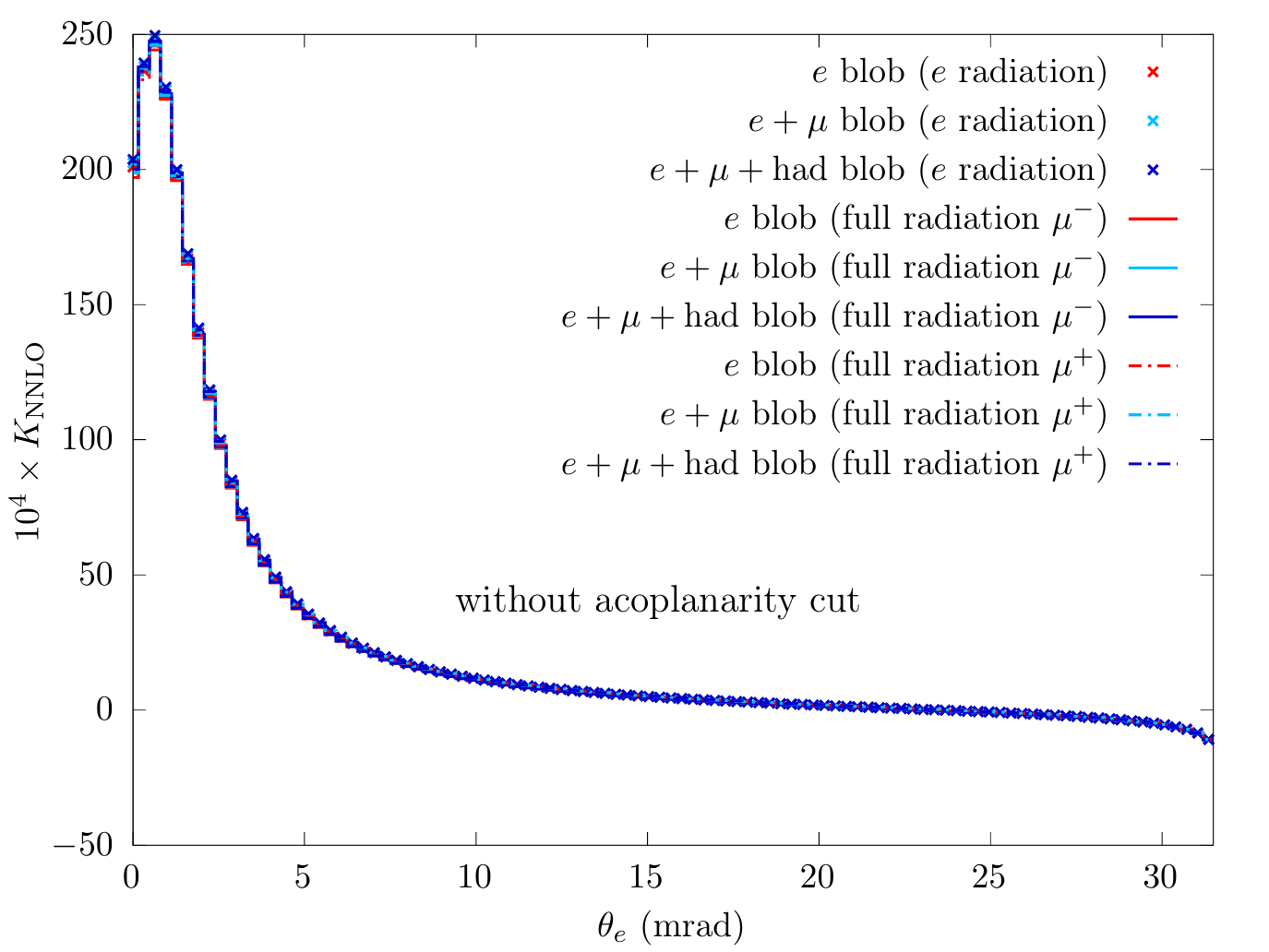}%
    \includegraphics[width=0.5\textwidth]{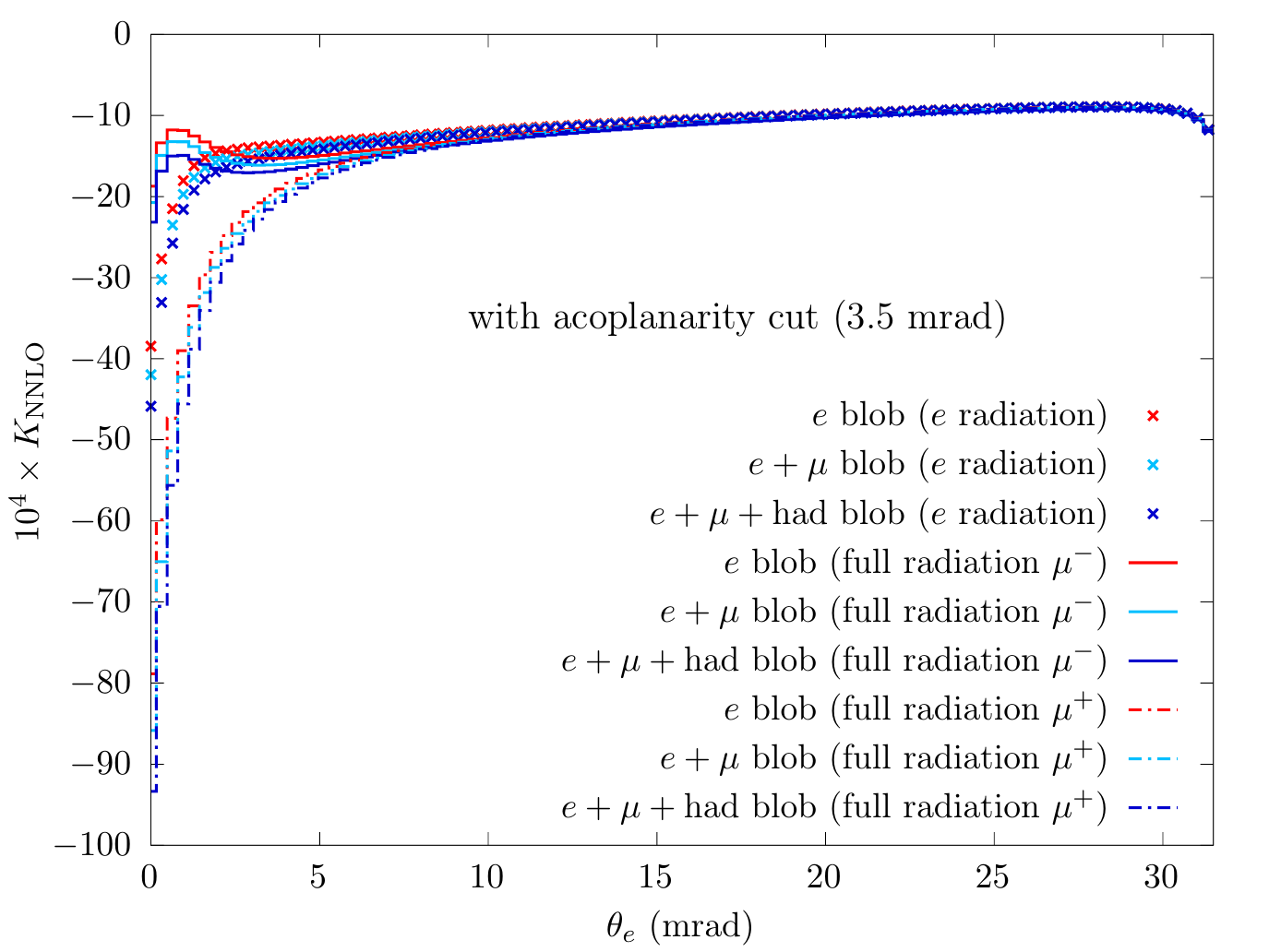}
  \caption{Left: differential $K_\text{NNLO}$ factor on $\theta_e$
    distribution, including the complete set of NNLO virtual leptonic
    pair corrections. Right: the same observable including the
    acoplanarity cut.  }
  \label{Fig:e_angle_vp_1L}
  \end{center}
\end{figure}
\begin{figure}[ht]
  \begin{center}
    \includegraphics[width=0.5\textwidth]{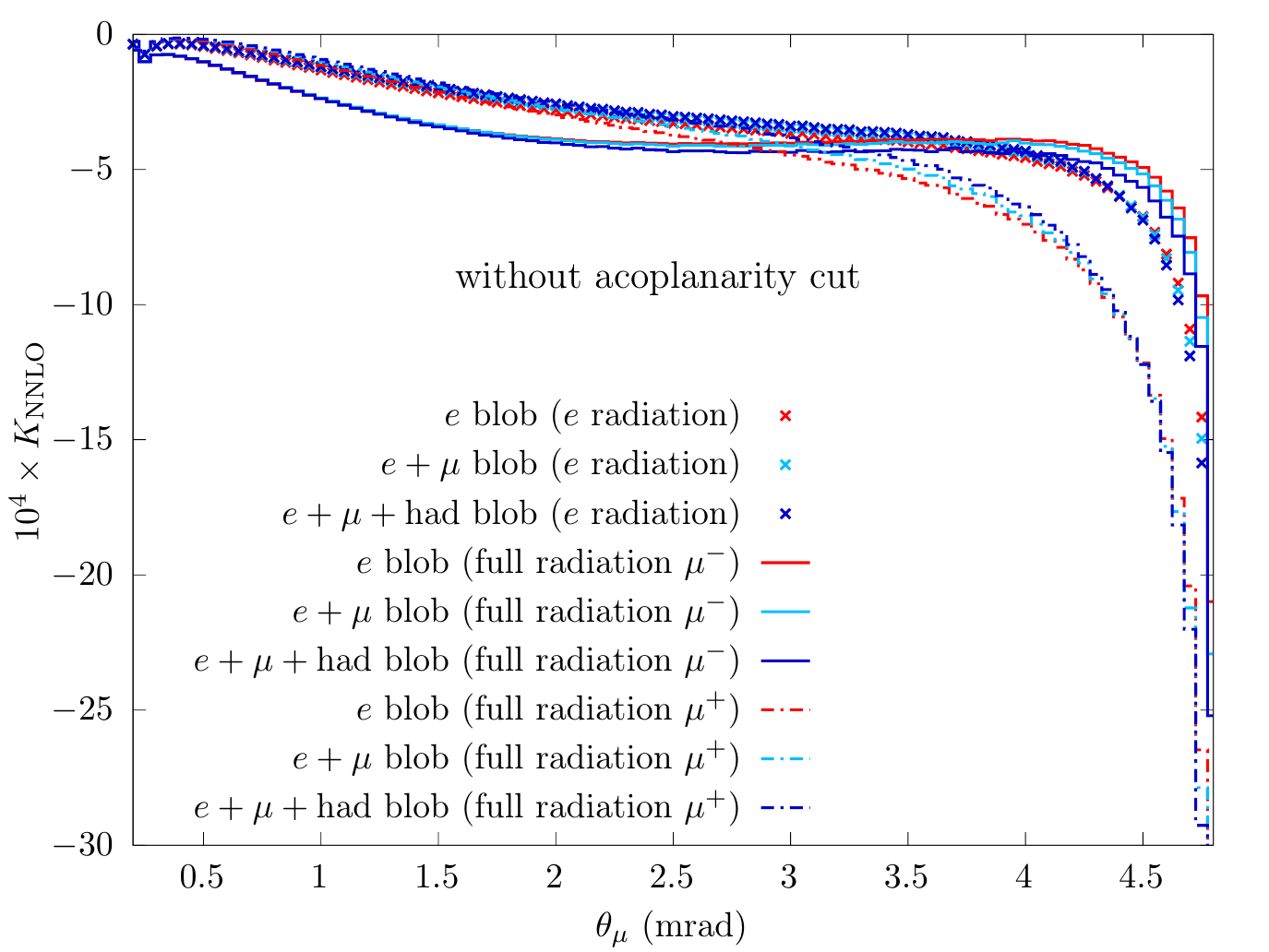}%
    \includegraphics[width=0.5\textwidth]{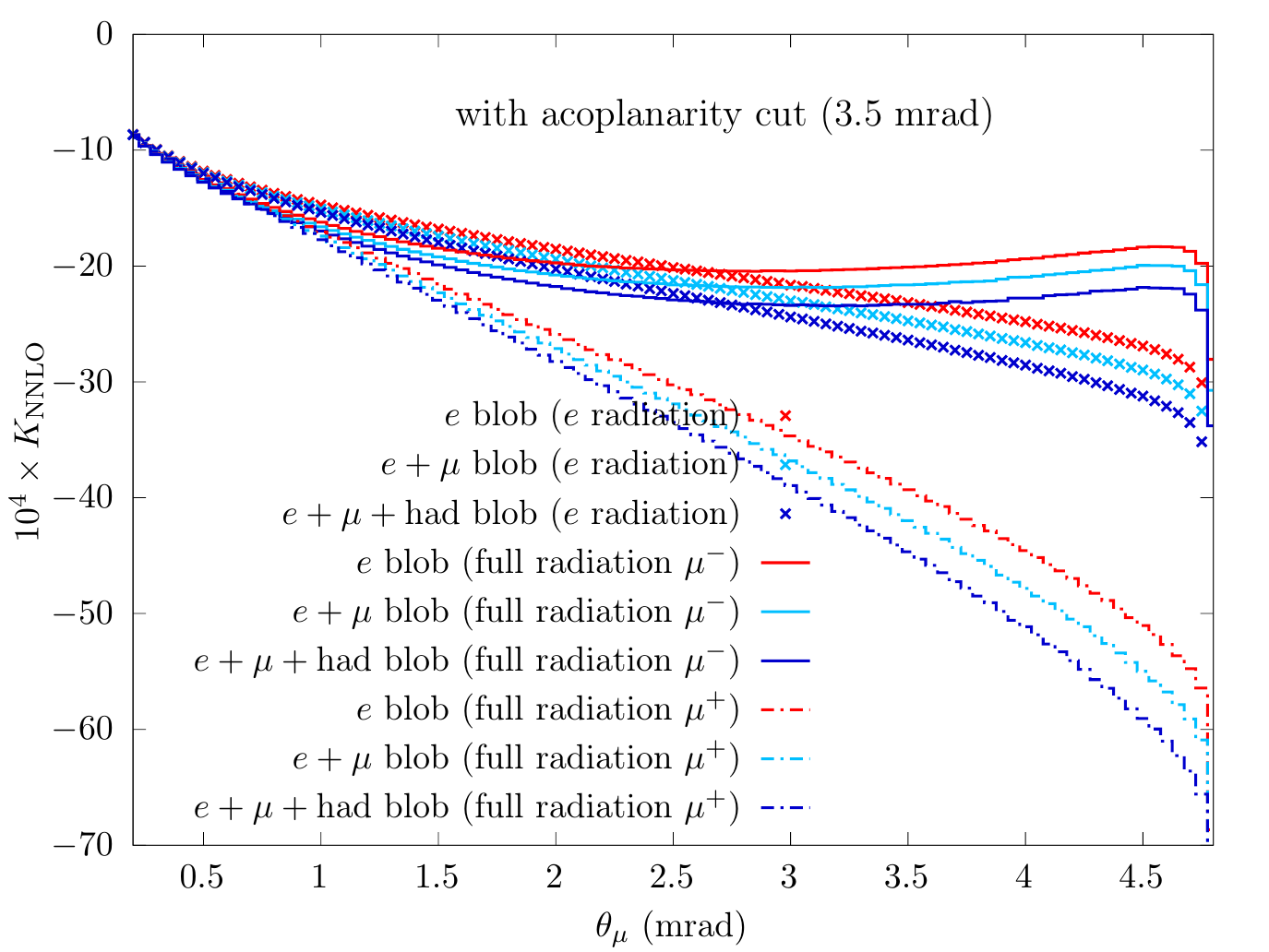}
  \caption{Left: differential $K_\text{NNLO}$ factor on $\theta_\mu$
    distribution, including the complete set of NNLO virtual leptonic
    pair corrections. Right: the same observable including the
    acoplanarity cut.  }
  \label{Fig:mu_angle_vp_1L}
  \end{center}
\end{figure}
\begin{figure}[ht]
  \begin{center}
    \includegraphics[width=0.5\textwidth]{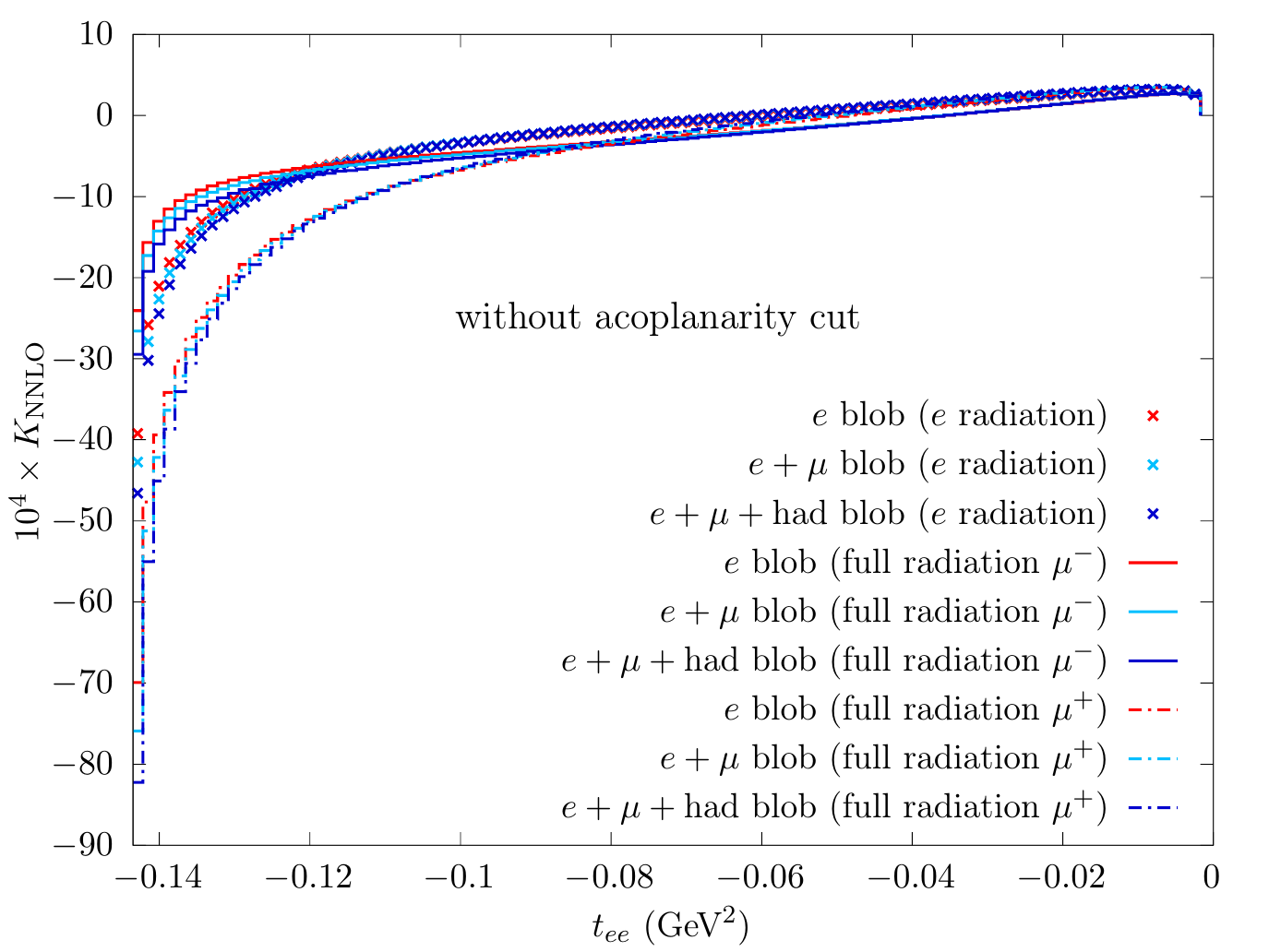}%
    \includegraphics[width=0.5\textwidth]{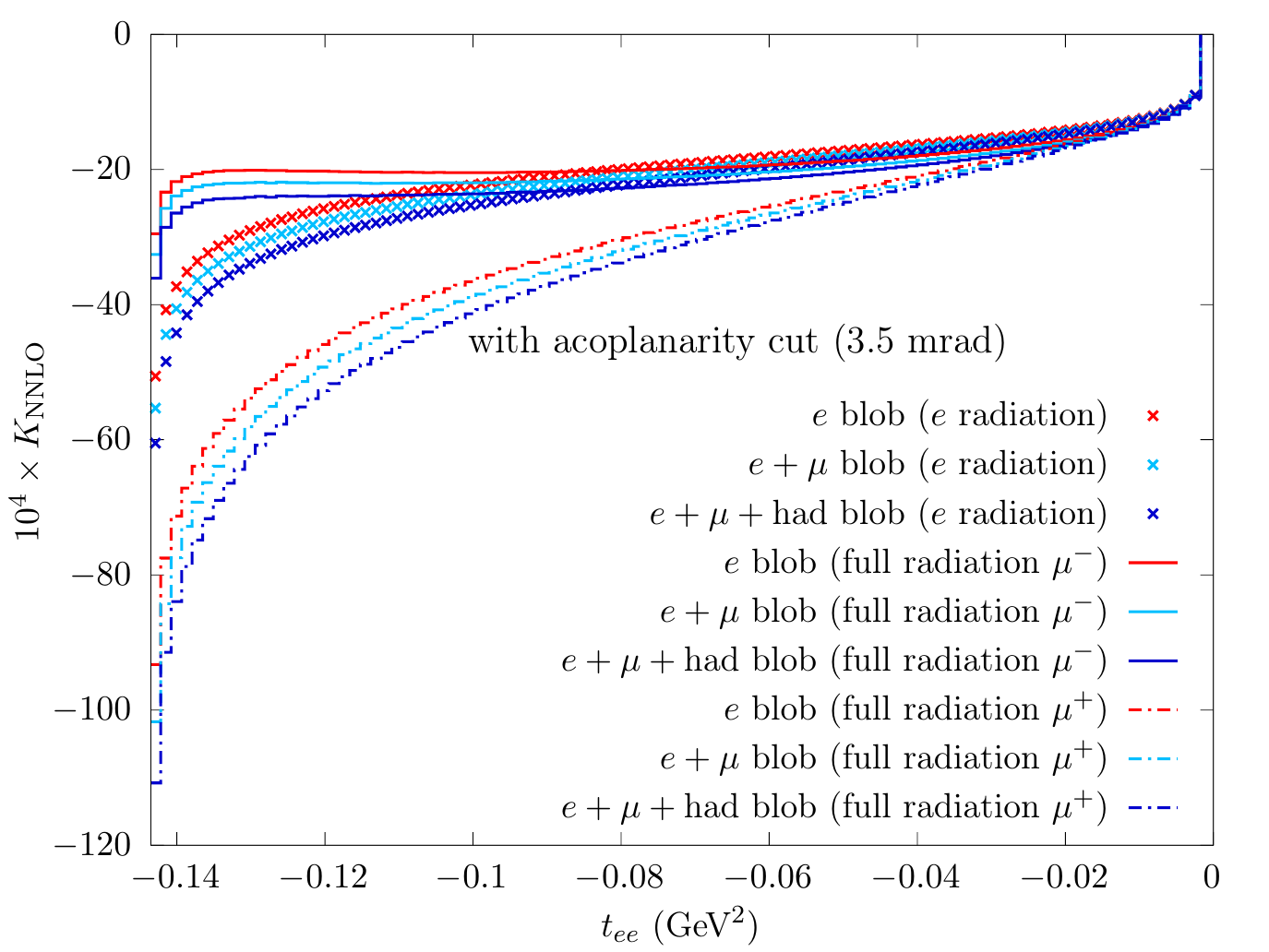}
  \caption{Left: differential $K_\text{NNLO}$ factor on $t_{ee}$
    distribution, including the complete set of NNLO virtual leptonic
    pair corrections. Right: the same observable including the
    acoplanarity cut.  }
  \label{Fig:t24_vp_1L}
  \end{center}
\end{figure}

Figure~\ref{Fig:e_angle_vp_1L} shows the impact of the virtual pair
corrections on the electron scattering angle distribution for
different levels of approximation: radiation from the electron leg
only or the complete calculation for $\mu^-$ or $\mu^+$ beam. For each
of these cases, three different contributions are considered: only
electron loop, electron plus muon loop and electron plus muon plus
hadronic loop. On the left, in the absence of any acoplanarity cut,
all the above described contributions are almost degenerate, with very
large corrections up to about $2.5\%$ for small scattering angles,
driven by the electron loop on top of radiation from the electron
line.  On the right, the additional acoplanarity cut changes the sign
of the corrections, which range from about $-0.1\%$ for large
scattering angles to several $-0.1\%$ for $\theta_e \to 0$. In
particular, the contribution from the electron line radiation reaches
the value of about $-0.4\%$, driven by the electron closed loop. The
muon and the hadronic loops are of the same order of a few
$0.01\%$. The full corrections (solid lines for $\mu^-$ beam and
dot-dashed line for $\mu^+$ beam) display a difference with respect to
the radiation from the electron line at the few 0.1\% level for small
scattering angles.

Figure~\ref{Fig:mu_angle_vp_1L} shows the impact of the virtual pair
corrections on the muon angle distribution~\footnote{In this case and
similarly thorough the rest of the paper, we do not show muon
scattering angles $\theta_\mu < 0.2\text{ mrad}$ because the
corrections are huge and would spoil the readability of the plots. The
reason for this feature is that the LO prediction becomes tiny in the
angular range below $0.2\text{ mrad}$, as a consequence of the applied
acceptance cuts, while 3-body (and later 4-body) events populate the
region, not being constrained by the $2\to 2$ kinematics.}. In this
case, the corrections grow to dozens of $0.01\%$ as $\theta_\mu$
reaches its maximum. We notice a flattening of the $K_\text{NNLO}$
factor when the acoplanarity cut is added (right). The effect is
generally dominated by electron loop insertion, and in the full
calculation corrections for an incident $\mu^+$ or $\mu^-$ are
opposite with respect to corrections on the electron line
only. Finally, figure~\ref{Fig:t24_vp_1L} shows the impact on the
momentum transfer $t_{ee}$, where similar considerations apply.

We conclude the section by remarking that the application of extra
elasticity cuts, in particular \texttt{cut 1+3}, does not change the
overall effects and size of the corrections considered here. In
contrast, they have a much more dramatic effect in reducing the
reducible background due to real electron pair emission, as will be
discussed in the following
section~\ref{sec:numerics-real-pairs-exclusive}.

\subsection{NNLO real pair corrections, asymmetric event selection (academic case)}
\label{sec:numerics-real-pairs-inclusive}
In this section we present the impact of the emission of a real
leptonic pair, treated in an inclusive way ({\it i.e.} with asymmetric
cuts). To exclude diagrams with the exchange of identical particles,
we consider in this section the processes $\mu^\pm e^- \to \mu^\pm e^-
\tau^+ \tau^-$, setting $m_\tau = m_e$, and adopt the generalised
event selection criteria described in the introduction of this
section. At the end of this subsection, for completeness, we consider
also the negligible contribution due to the emission of an extra muon
pair ($m_\tau = m_\mu$).

\begin{figure}[ht]
  \begin{center}
    \includegraphics[width=0.5\textwidth]{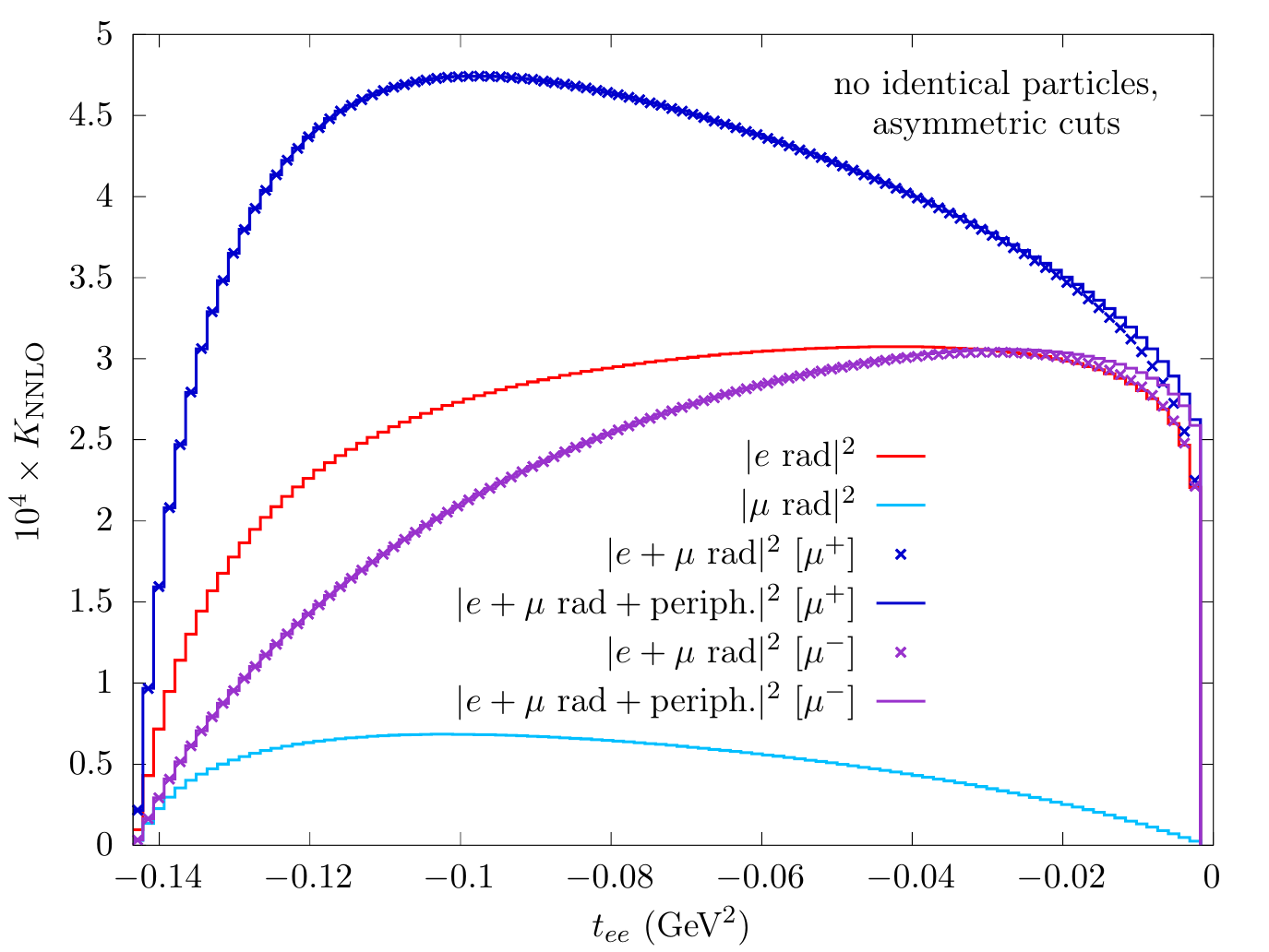}%
    \includegraphics[width=0.5\textwidth]{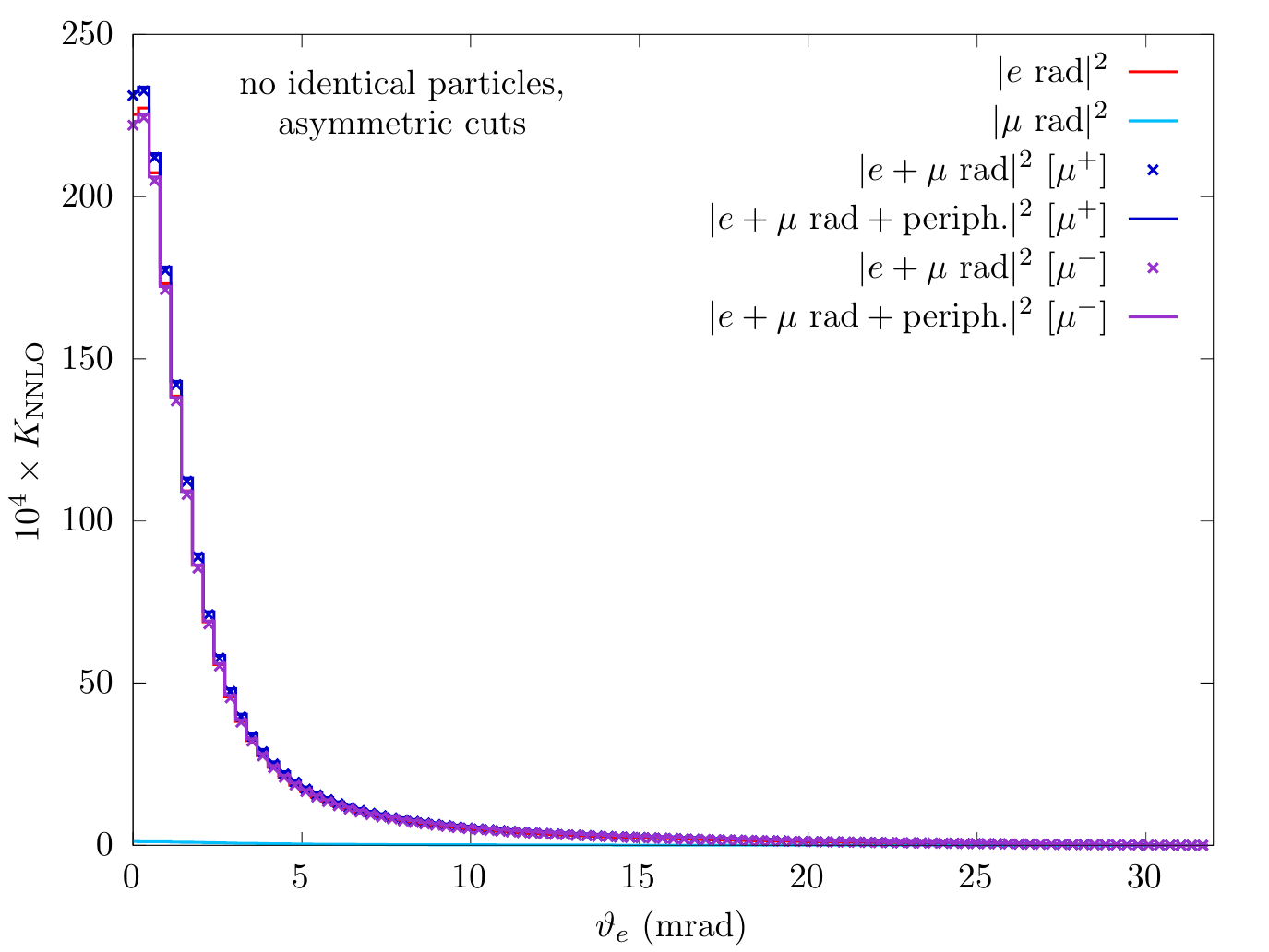}
  \caption{Real $e^+ e^-$ radiation effects with academic event
    selection on $t_{ee}$ (left) and $\theta_{e}$ (right)
    distribution.}
\label{Fig:t24-real-academic}
  \end{center}
\end{figure}

In the left panel of figure~\ref{Fig:t24-real-academic} we show the
contribution of the ``$\tau$'' pair (with $m_\tau = m_e$) on the
$t_{ee}$ distribution. Three different gauge invariant classes of
diagrams are considered: radiation from the electron line only (red
line), radiation from the muon line only (light-blue line), radiation
from both electron and muon lines (see
figure~\ref{Fig:real-from-electron}~(a)-(d)), including their
interferences but neglecting the peripheral diagrams
(figure~\ref{Fig:real-from-electron}~(e) and (f)).  For the latter two
options, the blue and violet crosses refer to the $\mu^+$ and $\mu^-$
beam cases, respectively. The results obtained with the complete
matrix elements, including also the peripheral diagrams, are displayed
with the blue and violet line for the $\mu^+$ and $\mu^-$ beam,
respectively. As a general comment we can see that the numerical
impact of the real $e^+ e^-$ radiation, on the $t_{ee}$ distribution,
is at the level of few $0.01\%$, with dominance of the radiation from
the electron line w.r.t. the radiation from the muon line. The
interference between the radiation from the electron and muon lines is
of different sign when flipping the charge of the incoming muon, as
already seen for the QED NLO corrections in
ref.~\cite{Alacevich:2018vez}. The effect of the peripheral diagrams
is almost negligible.
\begin{figure}[ht]
  \begin{center}
    \includegraphics[width=0.5\textwidth]{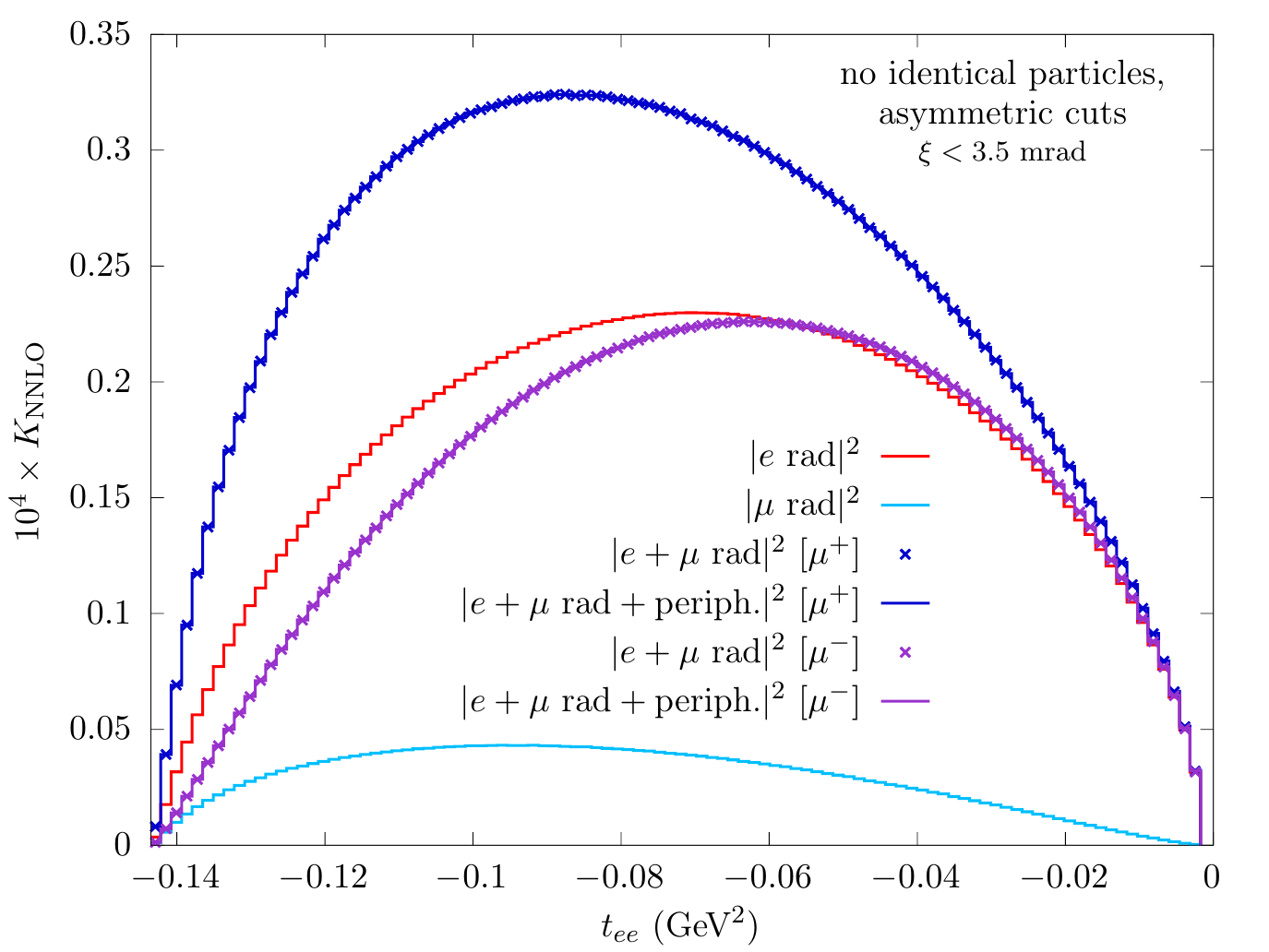}%
    \includegraphics[width=0.5\textwidth]{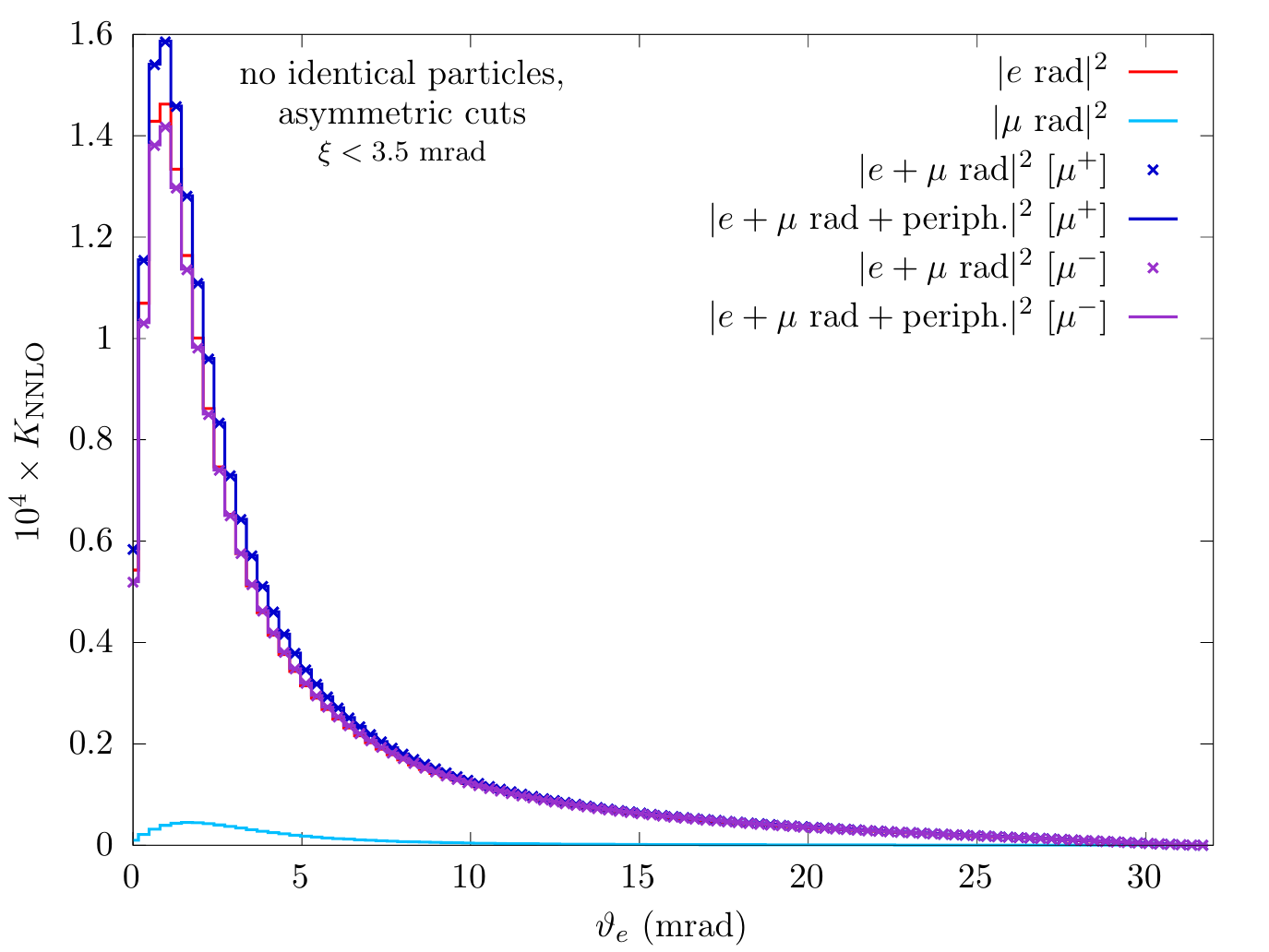}
  \caption{Real $e^+ e^-$ radiation effects with academic event
    selection on $t_{ee}$ (left) and $\theta_e$ (right) distribution,
    in the presence of the acoplanarity cut.  }
\label{Fig:ethtlab-real-acocut-academic}
  \end{center}
\end{figure}

The right panel of figure~\ref{Fig:t24-real-academic} shows the
electron scattering angle distribution, normalised to the LO prediction, where the
large correction factor for the radiation from the electron line is
clearly visible for $\theta_e \to 0$, analogously to what happens for
the NLO QED correction. As commented in ref.~\cite{Alacevich:2018vez},
the effect is in part due to the vanishing of the LO differential
cross section as $\theta_e \to 0$. For this reason, in
ref.~\cite{Alacevich:2018vez} the acoplanarity cut has been
introduced, which allows to partially remove the enhancements.
Applying the same acoplanarity cut on our signature, for the $t_{ee}$
observable we obtain the results displayed in the left panel of
figure~\ref{Fig:ethtlab-real-acocut-academic}, where the maximum of
the correction is reduced to about $0.033\%$, for the case of an
incoming $\mu^+$ beam.  The corresponding corrections on the
$\theta_e$ observable are shown in the left panel of figure
\ref{Fig:ethtlab-real-acocut-academic}, and also here we remark the
large suppression factor induced by the acoplanarity cut.

Within this case, where we academically suppose one final state
electron to be distinguishable from those of the extra pairs, the
contribution from the peripheral diagrams is almost negligible for all
the considered observables.

We also notice that on the $K_\text{NNLO}$ factor for the $t_{ee}$
distribution, in the most inclusive case, without the acoplanarity
cut, corrections due to the emission of an extra $e^+e^-$ pair are
positive and of the same order of magnitude as the negative
corrections due to VP insertion on photonic vertex corrections,
discussed in section~\ref{sec:numerics-vp-on-vertex}. We see indeed a
partial cancellation between real and virtual pair corrections, for
this subset of contributions.

Before considering also identical particles and fully realistic event
selection on real pair emission, we show in this academic case the
impact of $\mu^\pm e^-\to\mu^\pm e^-\mu^+\mu^-$ events ({\it i.e.}
$\mu^\pm e^-\to\mu^\pm e^-\tau^+\tau^-$ with $m_\tau = m_\mu$) on the
$t_{ee}$ distribution, for completeness.
\begin{figure}[ht]
  \begin{center}
    \includegraphics[width=0.6\textwidth]{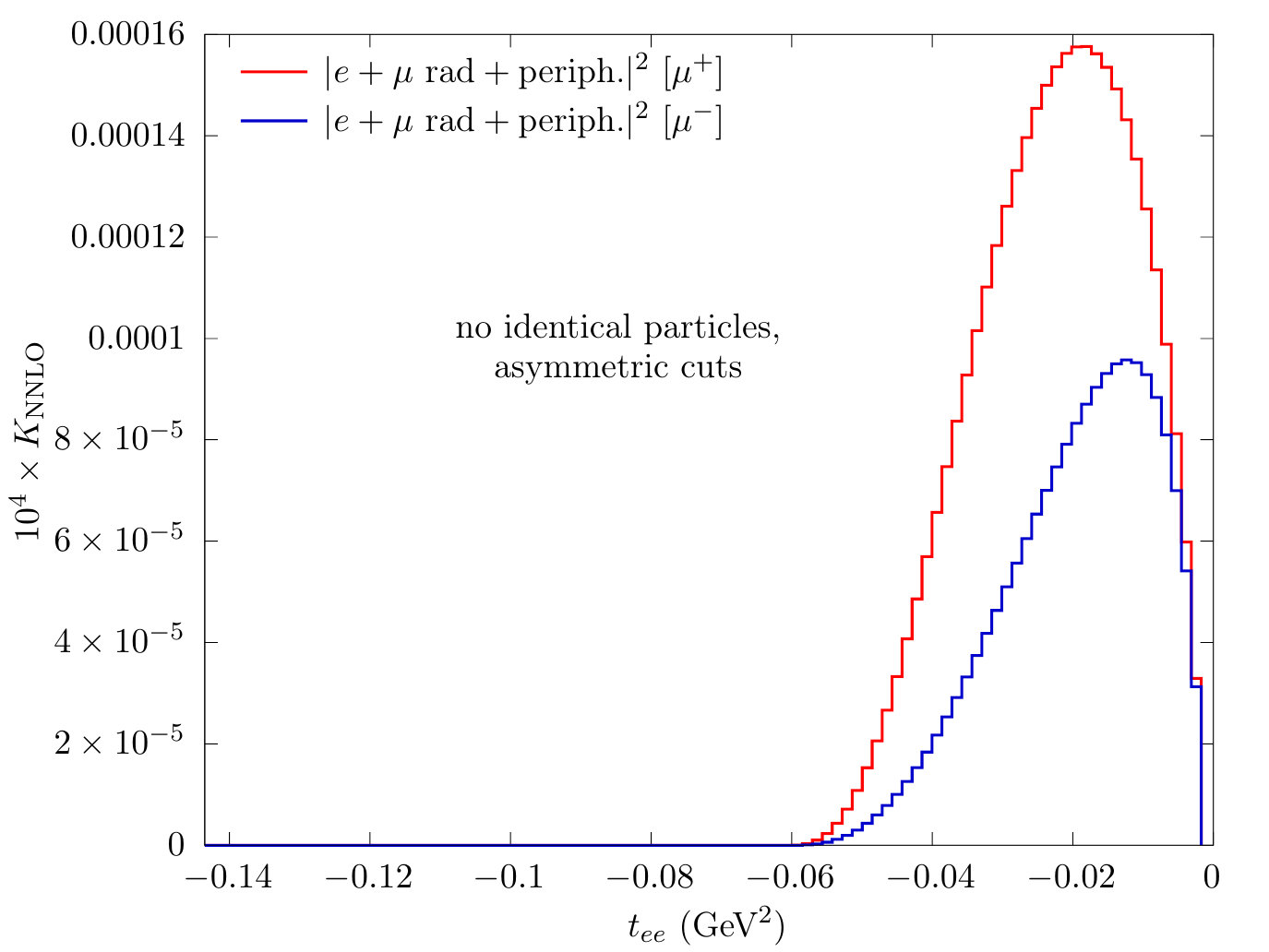}
  \caption{
  Impact of the $\mu^\pm e^-\to\mu^\pm e^-\mu^+\mu^-$ process on the $t_{ee}$
  distribution, for the academic selection cut.
  }
  \label{Fig:muonpairsacademic}
  \end{center}
\end{figure}
It is clear from figure~\ref{Fig:muonpairsacademic}, where the red
curve shows the incoming $\mu^+$ case and the blue one the incoming
$\mu^-$ case, that this process gives a totally negligible
contribution, reaching at maximum a few units in $10^{-8}$ w.r.t. the
LO differential cross section. We also notice that below $t_{ee}\simeq
-0.06\text{ GeV}^2$ there are no events because of phase space
constraints~\footnote{On the same grounds, we expect charged pion pair
emission ($\mu^\pm e^-\to\mu^\pm e^-\pi^+\pi^-$) to be even more
suppressed due to the further reduction of available phase space
because of the larger pion mass.}.

\subsection{NNLO real pair corrections, realistic event selection}
\label{sec:numerics-real-pairs-exclusive}
The discussion on real electron pair radiation of
section~\ref{sec:numerics-real-pairs-inclusive} shows the relevance of
real pair emission in the NNLO complete calculation. However, those
results are purely academic since the analysis is completely inclusive
on the emitted electron pair and does not take into account the MUonE
realistic event selection, where the four final state leptons can
produce up to four tracks in the detector. According to the discussion
of section~\ref{sec:numerics}, to mimic an elastic event, we can
require that only two tracks (one muon-like, one electron-like) are
seen in the detector. In this way, without charge identification, in
principle every pair out of the six possible pairings can hit the
detector while the remaining two particles are below energy threshold
and/or out of the angular allowed range. This requires the use of the
full matrix elements for the calculation of the cross sections of the
processes $\mu^\pm e^- \to \mu^\pm e^- e^+ e^-$ (with MUonE realistic
event selection we are considering here, no events for the process
$\mu^\pm e^- \to \mu^\pm e^- \mu^+ \mu^-$ satisfy the criteria,
because at least three charged visible particles are always present),
including all the diagrams with the two indistinguishable electrons
interchanged, which were neglected in the previous subsection, and the
adoption of symmetric cuts that do not distinguish identical particles
in the final state.

\begin{figure}[ht]
  \begin{center}
    \includegraphics[width=0.5\textwidth]{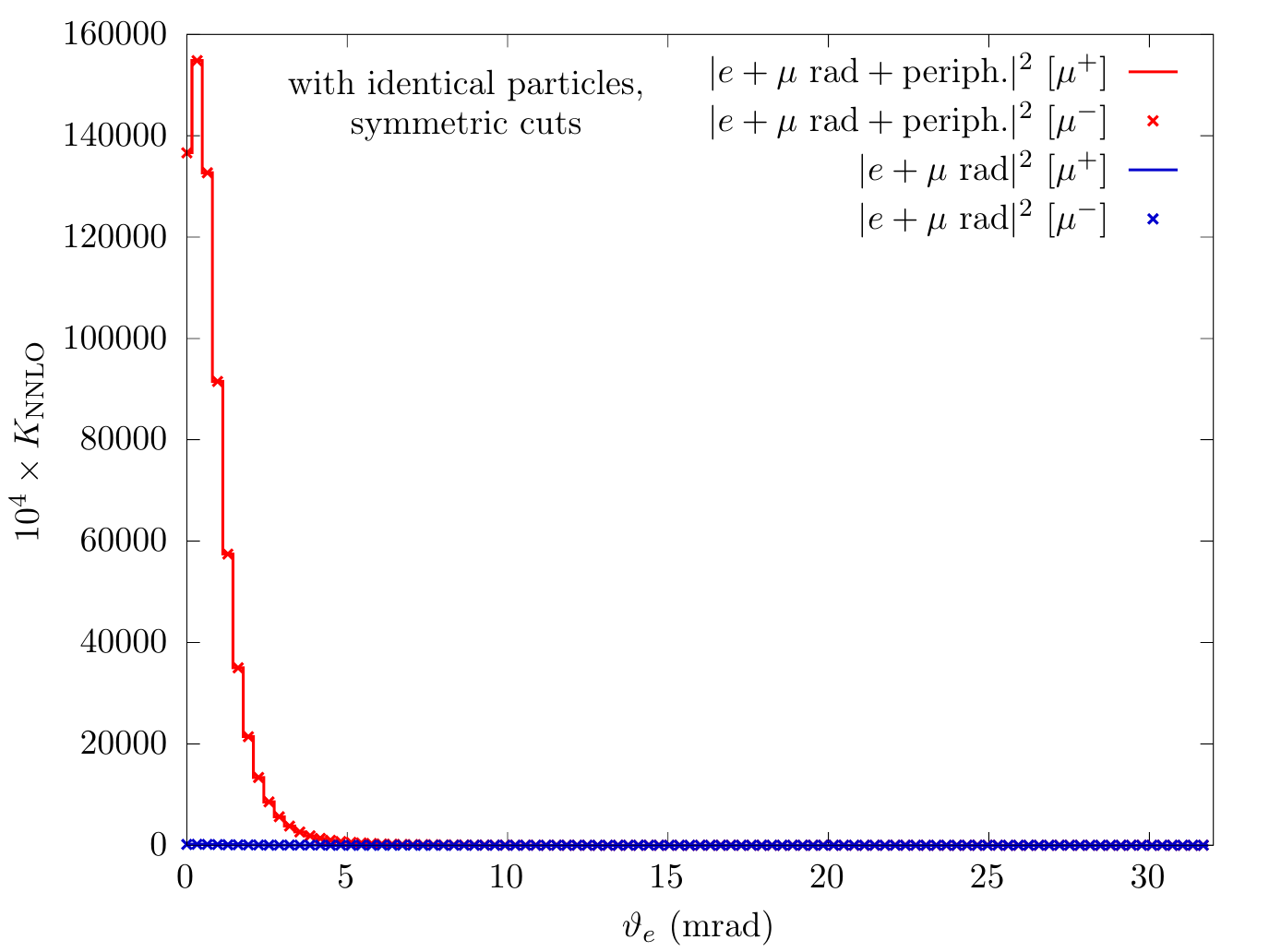}%
    \includegraphics[width=0.5\textwidth]{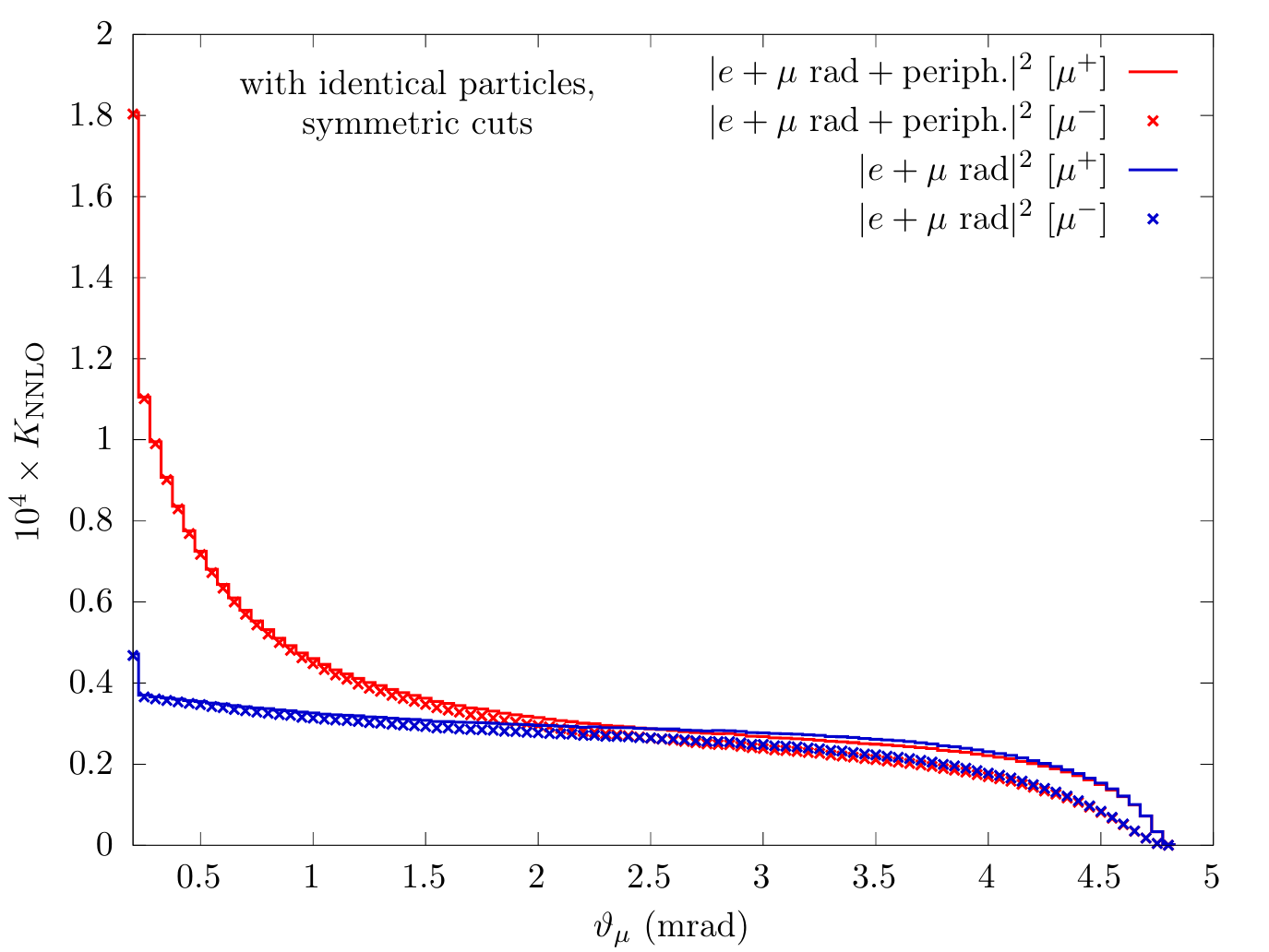}
  \caption{Differential $K_\text{NNLO}$ factor for $\theta_e$ (left)
    and $\theta_\mu$ (right) distributions for real $e^+ e^-$
    radiation with realistic event selection, without acoplanarity
    cut. $\theta_\mu$ is plotted above $0.2\text{ mrad}$ for
    readability.}
\label{Fig:emthlab-real-realistic}
  \end{center}
\end{figure}

This introduces a new important feature: there is no distinction
between the electron of the underlying tree-level and the electron or
positron of the emitted pair. As a consequence, in the photon
propagators the exchanged momentum can become very close to zero,
giving a dramatic enhancement in the cross section and in the
distributions, especially at the corner of the phase space. In this
enhancement the peripheral diagrams
(figures~\ref{Fig:real-from-electron}~(e)-(f)) give by far the largest
contribution. For instance, a sharp peak in a very small window of the
scattering angles can be seen in
figure~\ref{Fig:emthlab-real-realistic}, where it is evident that
``switching-on'' peripheral diagrams produces a huge effect.

\begin{figure}[ht]
  \begin{center}
    \includegraphics[width=0.5\textwidth]{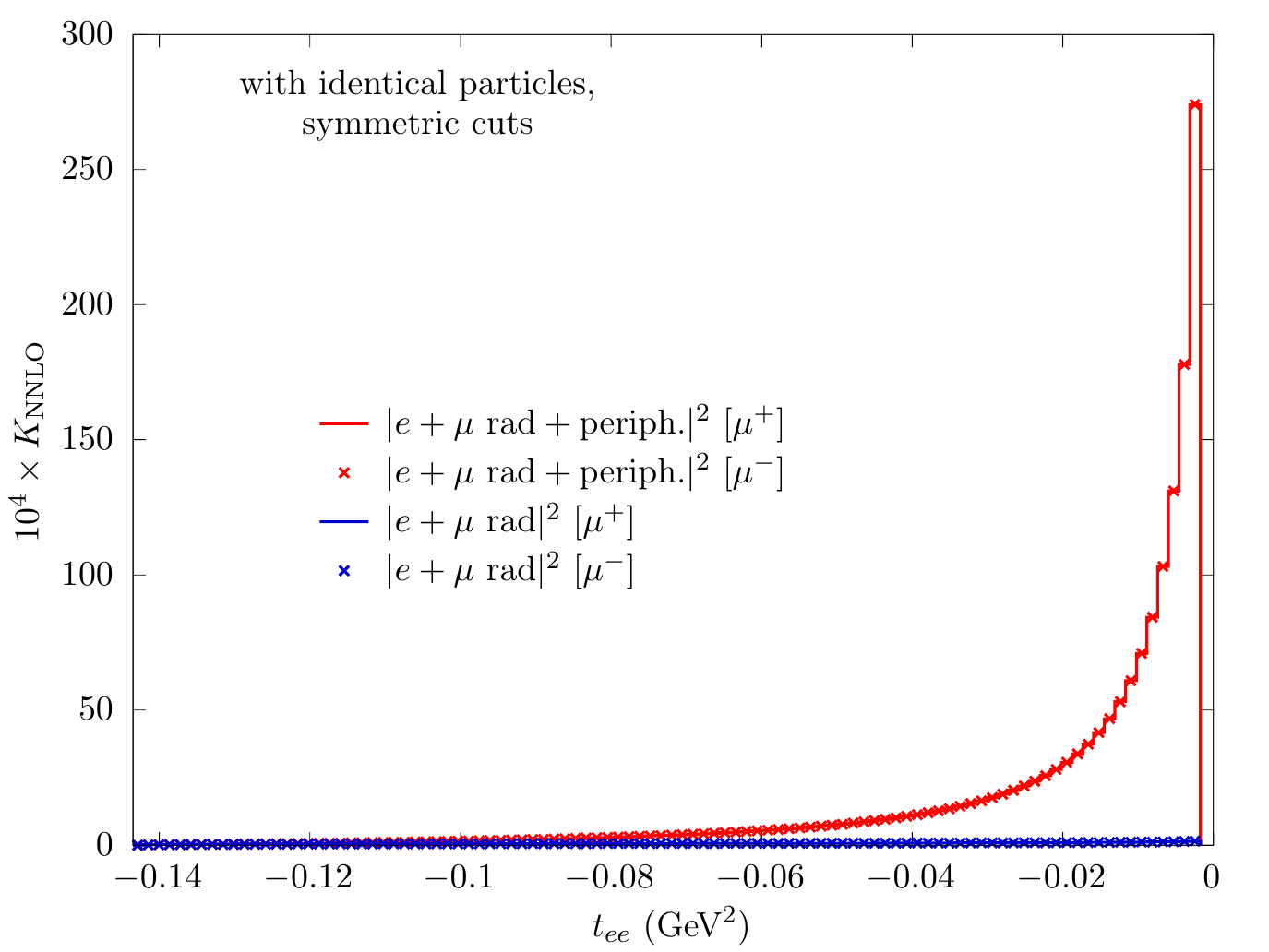}~
    \includegraphics[width=0.5\textwidth]{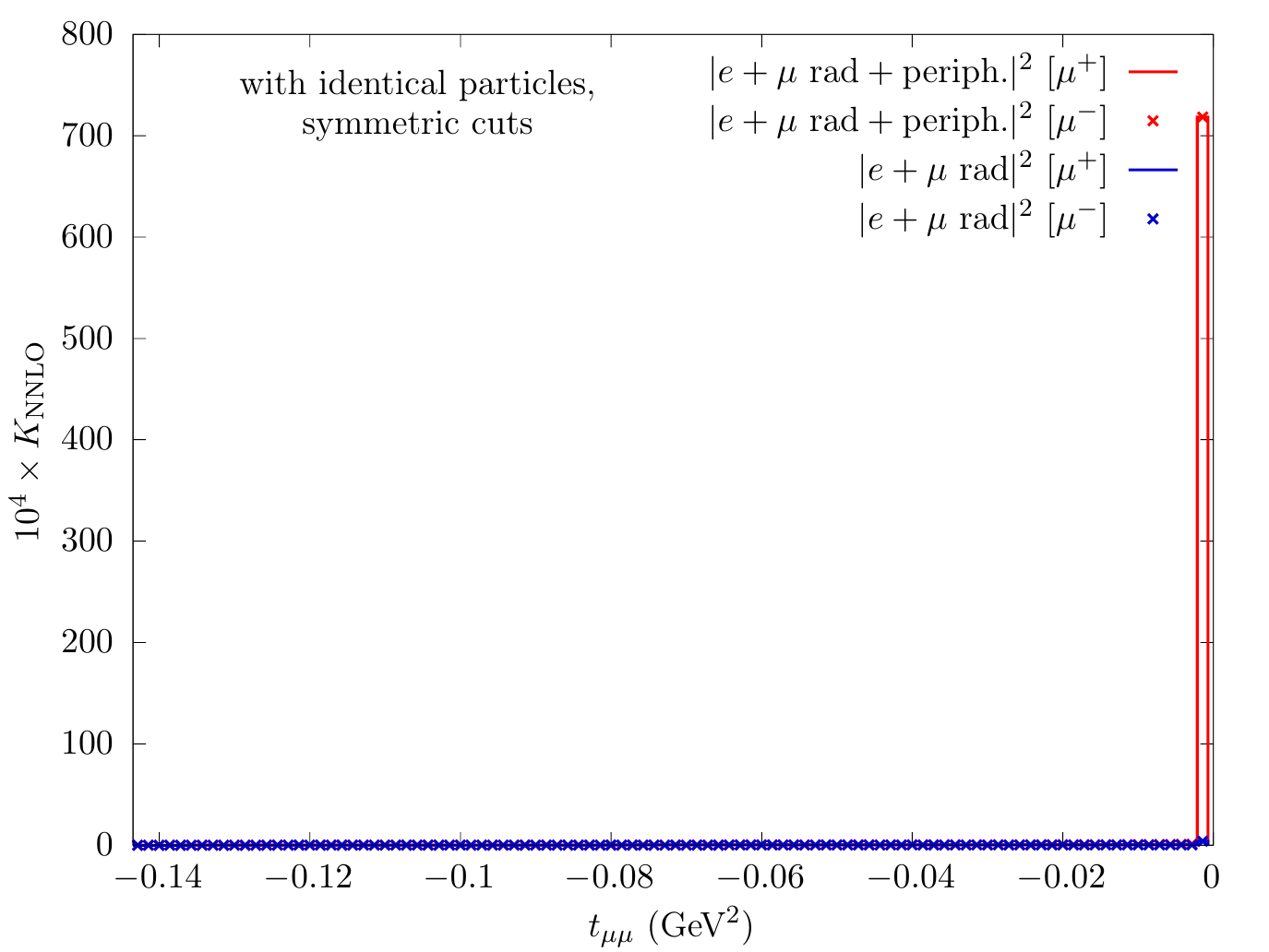}    
  \caption{Left: $t_{ee}$ differential $K_\text{NNLO}$ factor for
    real $e^+ e^-$ radiation with realistic event selection, without
    acoplanarity cut. Right: the same for $t_{\mu\mu}$.}
    \label{Fig:t24-real-realistic}
  \end{center}
\end{figure}

\begin{figure}[ht]
  \begin{center}
    \includegraphics[width=0.5\textwidth]{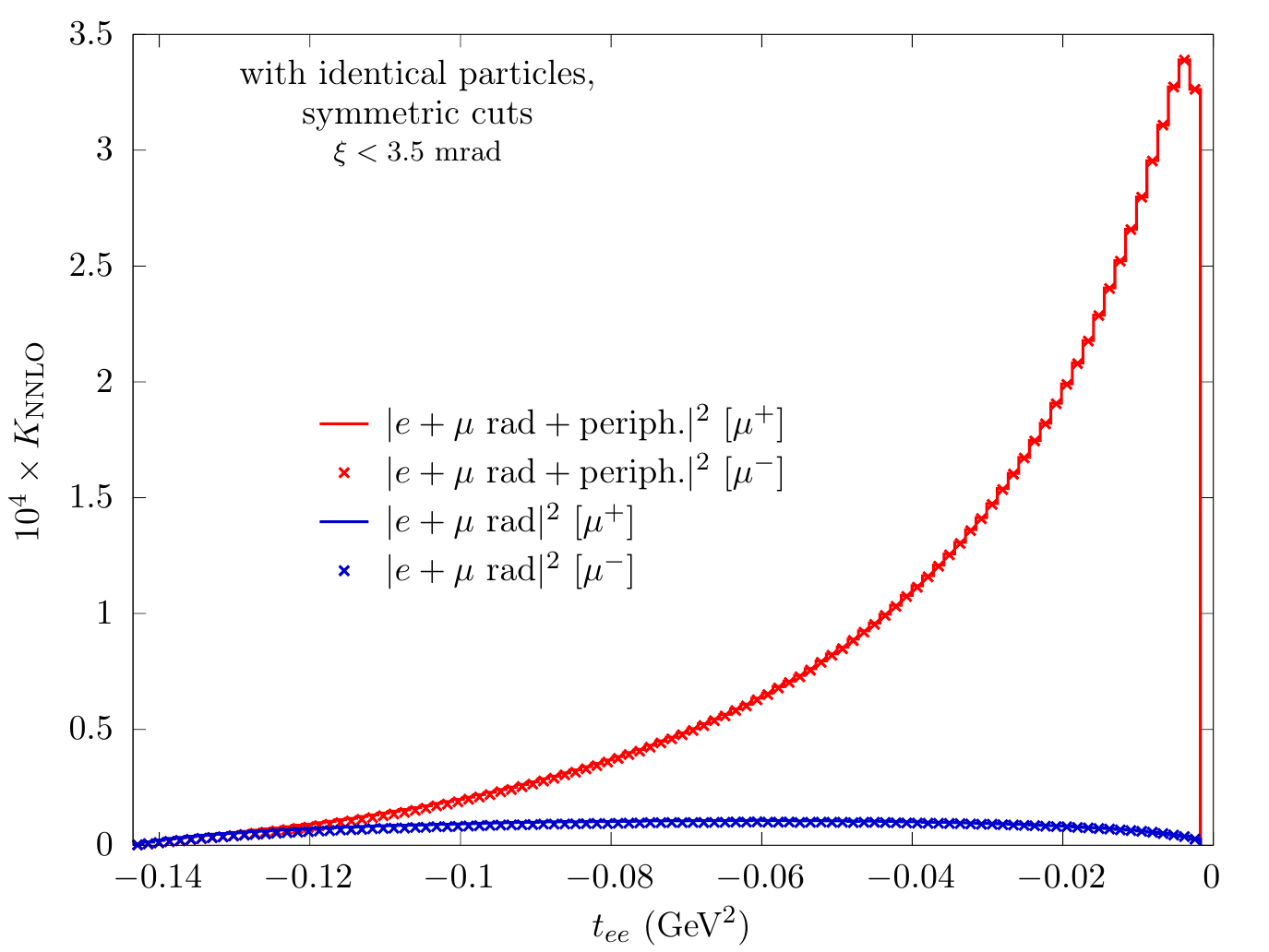}%
    \includegraphics[width=0.5\textwidth]{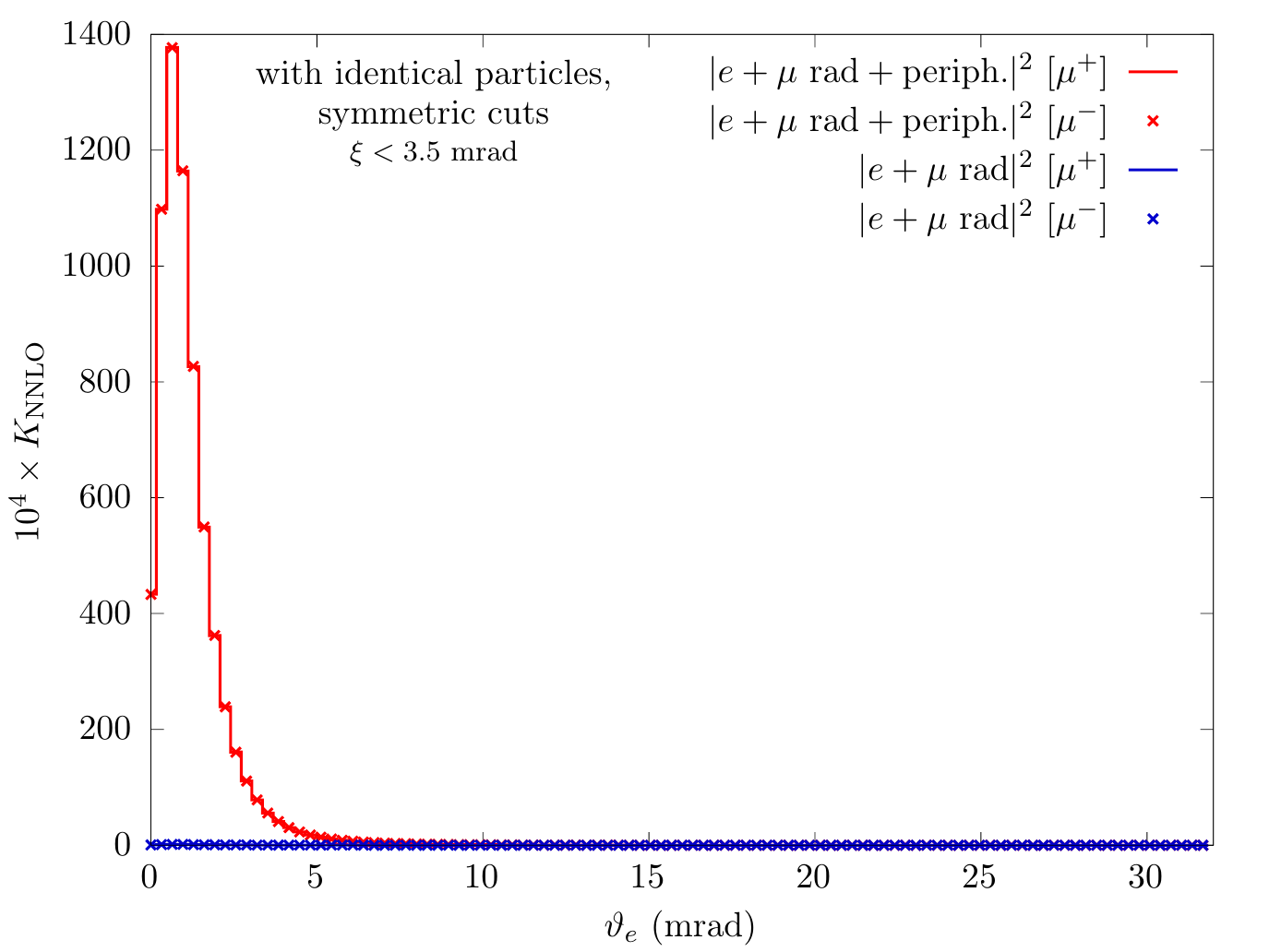}
  \caption{Left: Differential $K_\text{NNLO}$ factor on $t_{ee}$
    distribution for real $e^+ e^-$ radiation with realistic event
    selection, with acoplanarity cut.  Right: the same for
    $\theta_e$.}
    \label{Fig:et24-real-realistic-acocut}
  \end{center}
\end{figure}

The kinematical configurations related to these enhancements
correspond typically to small momentum transfer for $t_{ee}$ or
$t_{\mu\mu}$ (figure~\ref{Fig:t24-real-realistic}) and could spoil the
sensitivity of MUonE to the hadronic component of the QED running
coupling constant by distorting in a significant way the differential
cross sections where the signal to be measured is more important.
Even with the inclusion of the acoplanarity cut, the numerical impact
of real pair radiation remains at the level of $0.1\%$, with a large
contribution for $\theta_e \to 0$ reaching the $10\%$ level, as can be
seen in figure~\ref{Fig:et24-real-realistic-acocut}.

\begin{figure}[ht]
  \begin{center}
    \includegraphics[width=0.9\textwidth]{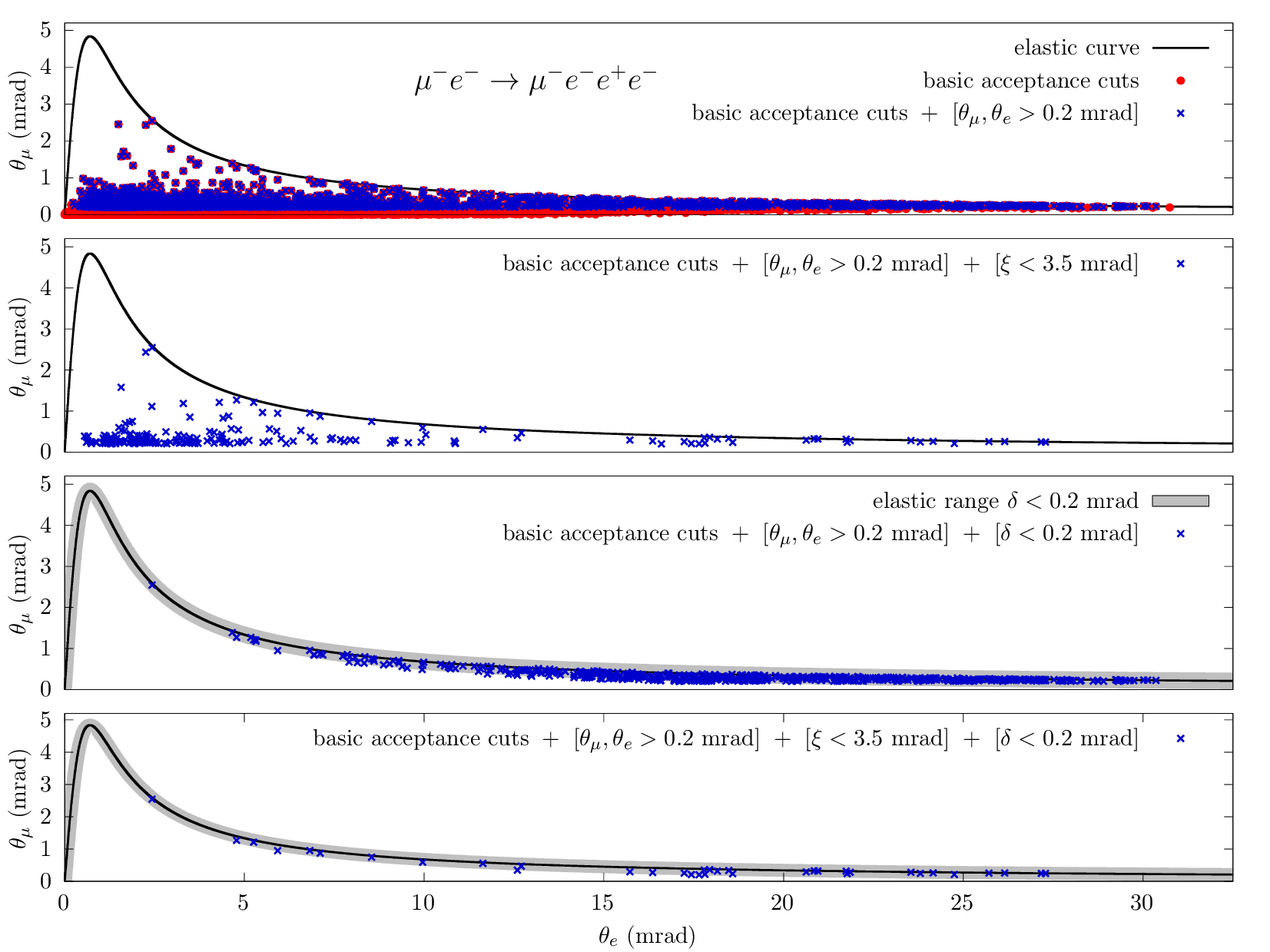}
  \caption{Scatter plot of $(\theta_e,\theta_\mu)$ points for $5\cdot
    10^5$ $\mu^-e^-\to\mu^- e^-e^+e^-$ simulated events.}
    \label{Fig:scatter-plot}
  \end{center}
\end{figure}

In order to reduce the impact of real pair radiation to a manageable
level, we propose to further filter events through the cuts labelled
as \texttt{cut 1-3} and described in the introduction to
section~\ref{sec:numerics}.  The effectiveness of the additional cuts
in suppressing real electron pair radiation events can be visualised
with the scatter plot of the $\theta_e$-$\theta_\mu$ angles
correlation displayed in figure~\ref{Fig:scatter-plot}.
The upper panel shows, with red points, the scatter distribution of
the coordinates $(\theta_{e},\theta_\mu)$ for a sample of $5\cdot
10^5$ events for the process $\mu^- e^- \to \mu^- e^- e^+ e^-$ with
only one muon-like and one electron-like track. It is clearly seen
that many points fall well outside the elastic correlation curve
(black line), in particular they gather in the region of small
$\theta_{e}$ and $\theta_\mu$.  The blue crosses display the same
sample after applying the additional cut
$\theta_{e},\theta_\mu >\theta_c$ (\texttt{cut 1}). It is clearly visible the removal of the
events with $\theta_\mu$ close to 0. The background rejection
efficiency of such a cut is of about $99.5\%$. In the second panel
from the top, the effect of adding the acoplanarity cut (\texttt{cut
  1+2}) is shown and the rejection efficiency increases to about
$99.96\%$. The third panel from the top displays the effect of
imposing the cut on the distance from the elasticity correlation
curve, $\delta < \delta_c$, on top of \texttt{cut 1}. In this case the
rejection efficiency is at the level of $99.89\%$.  Finally, the
bottom panel shows the effects of combining the three cuts
(\texttt{cut 1+2+3}). As can be seen, this combination is particularly
effective in reducing the real electron pair production, since only a
tiny fraction (as small as $0.007\%$) of background events passes
through the selection. It is interesting to observe the
complementarity between the $\xi$ and $\delta$ cuts. The latter alone
suppresses hard radiation contributions while allowing for a certain
degree of acoplanarity between the azimuthal angles of the two final
state tracks, due to soft transverse radiation.

\begin{figure}[ht]
  \begin{center}
    \includegraphics[width=0.5\textwidth]{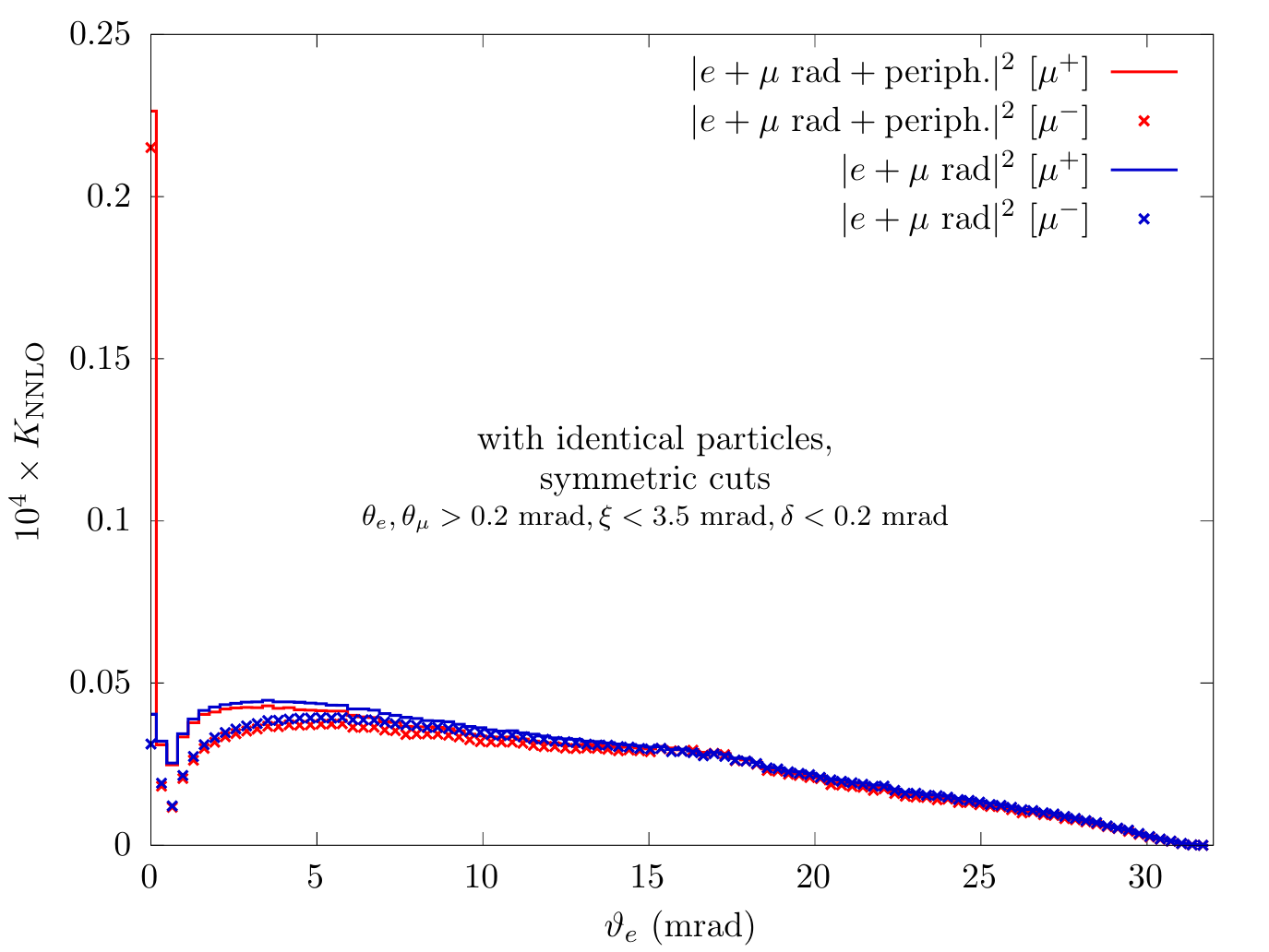}%
    \includegraphics[width=0.5\textwidth]{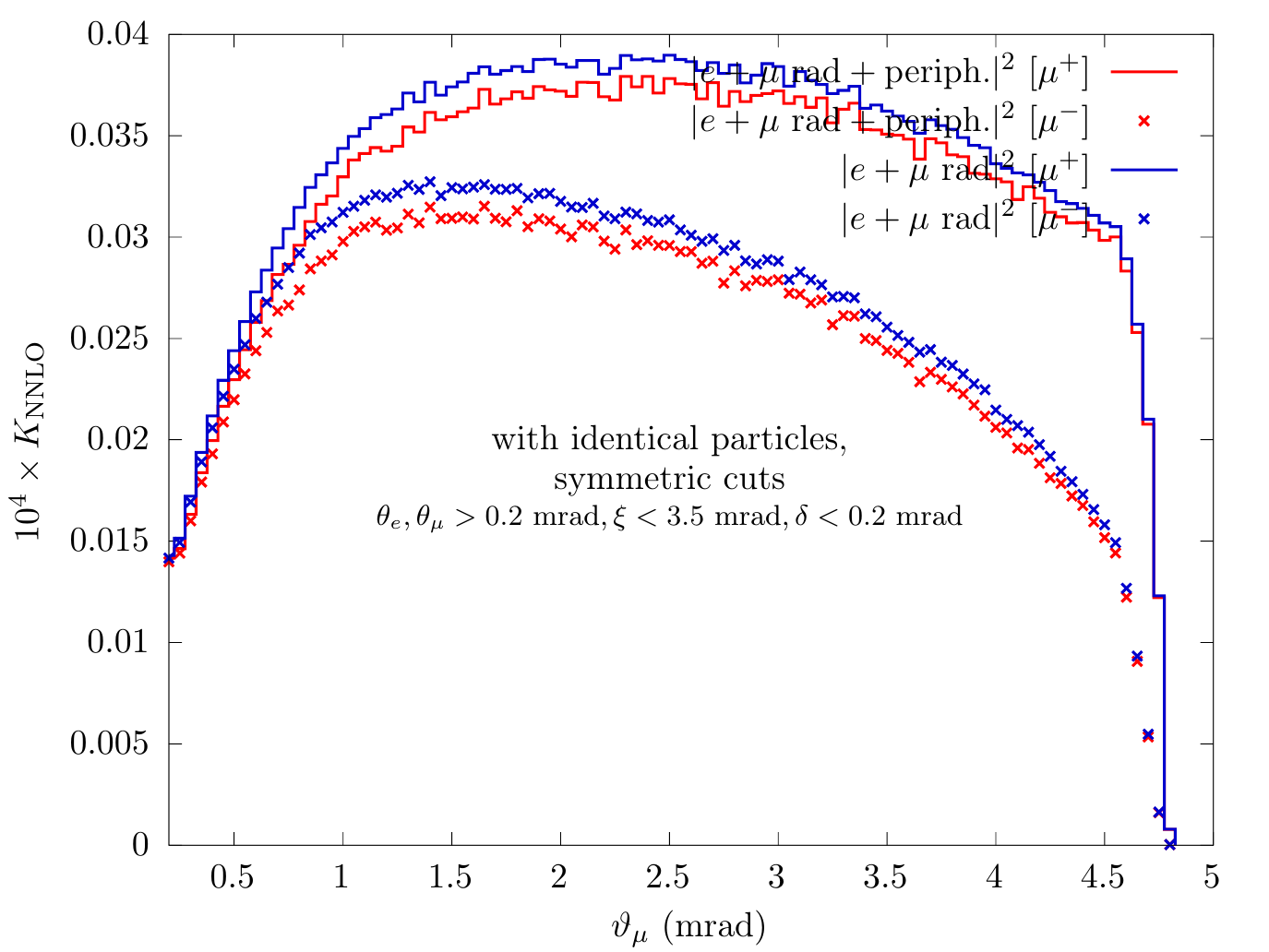}
  \caption{Differential $K_\text{NNLO}$ factor for $\theta_e$ (left)
    and $\theta_\mu$ (right) distributions for real $e^+ e^-$
    radiation with realistic event selection, with all cuts applied.}
\label{Fig:emthlab-real-realistic-allcuts}
  \end{center}
\end{figure}

\begin{figure}[ht]
  \begin{center}
    \includegraphics[width=0.5\textwidth]{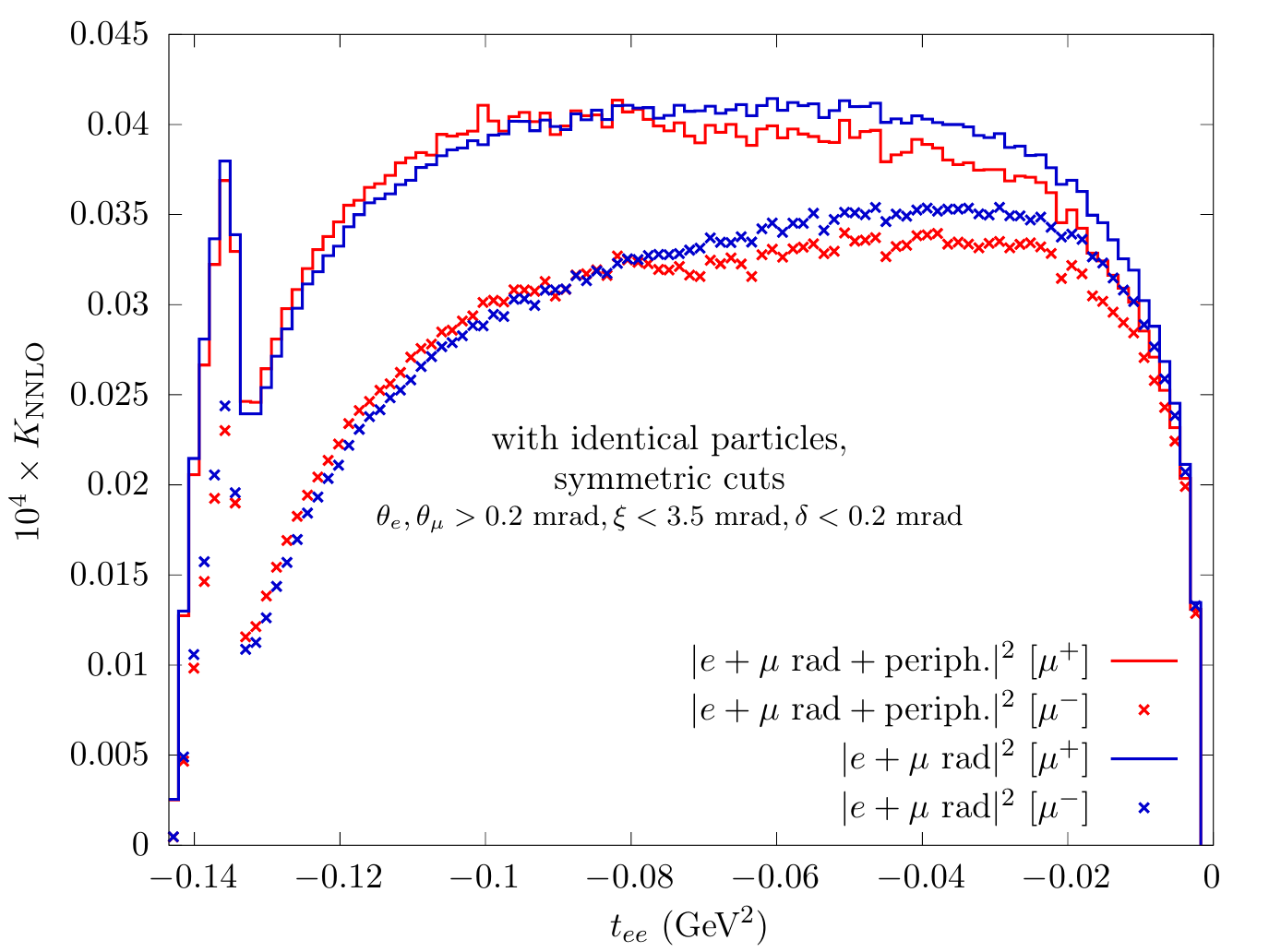}~
    \includegraphics[width=0.5\textwidth]{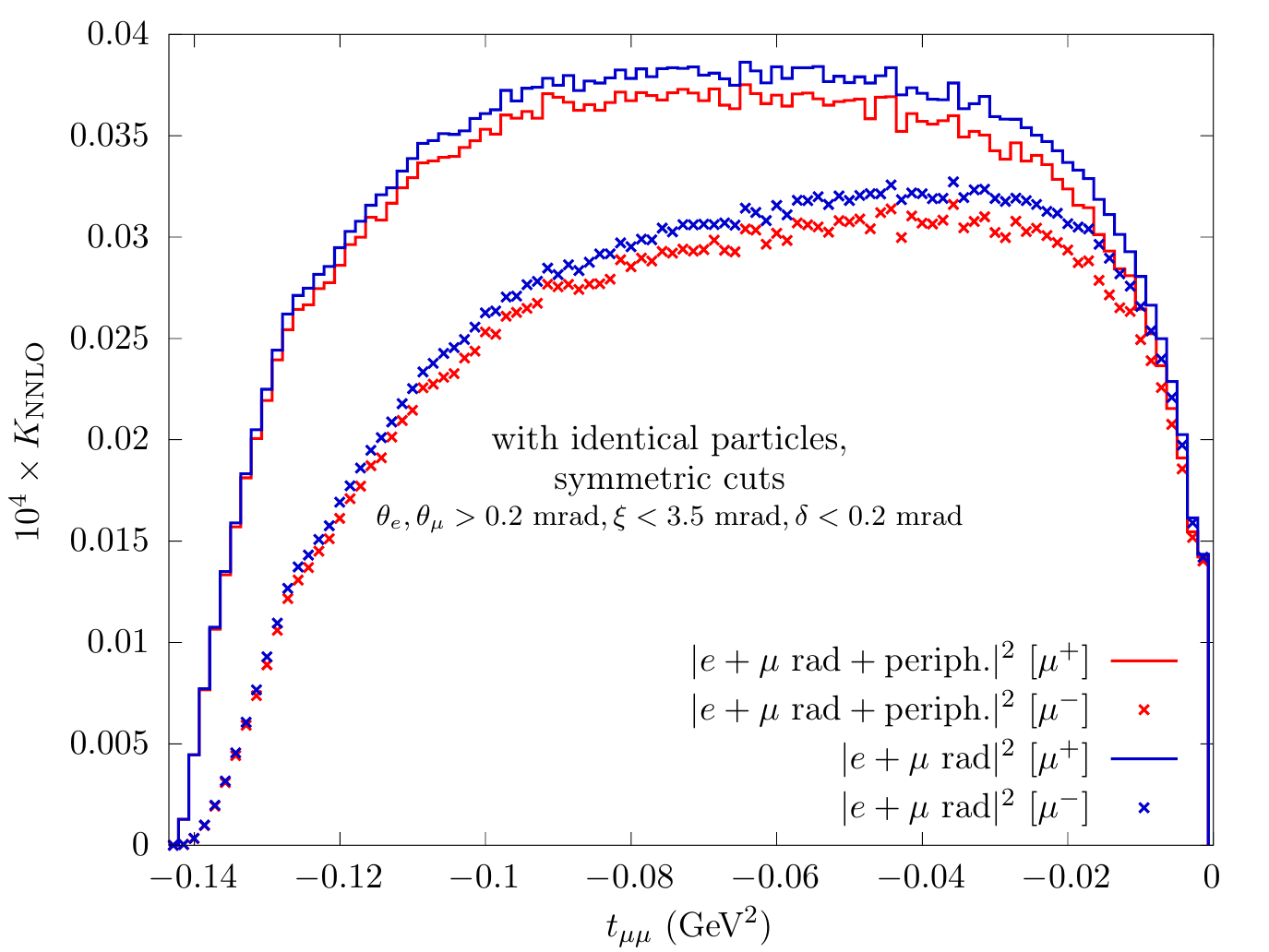}
  \caption{Left: Differential $K_\text{NNLO}$ factor on $t_{ee}$
    distribution for real $e^+ e^-$ radiation with realistic event
    selection, with all cuts applied. Right: the same for
    $t_{\mu\mu}$.}
    \label{Fig:t24-real-realistic-allcuts}
  \end{center}
\end{figure}

This analysis allows us to propose an improved event selection,
\texttt{cut 1-3}, as defined in section~\ref{sec:numerics} for the
analysis of MUonE data. As can be seen in
figures~\ref{Fig:emthlab-real-realistic-allcuts}
and~\ref{Fig:t24-real-realistic-allcuts}, with this event selection
the real $e^+e^-$ emission contribution gets reduced to the ppm level
over the full kinematical range. It is worth noticing, for instance in
the $\theta_e$ $K_\text{NNLO}$ factor, a peak on the leftmost bin: it
can be traced back to few events that, after applying \texttt{cut
  1+2}, lie in the region $\theta_e < \delta_c$ allowed by the
elasticity cut.

It is understood that more complete simulations will be needed to
select the optimum values for the cut parameters $\theta_c$,
$\delta_c$ and $\xi_c$. Here, we have only demonstrated that by a
sensible choice of the cut parameters the potentially large background
due to electron-pair emission can be reduced at the $10^{-5}$ level.

\section{Summary and prospects}
\label{sec:summary}
In this work we have discussed the calculation of the NNLO leptonic
corrections to $\mu^\pm e^-\to \mu^\pm e^-$ processes and their
implementation into the Monte Carlo code \textsc{Mesmer}, used for
simulations for the MUonE experimental proposal. Such a project aims
at a high precision measurement of the effective QED coupling constant
in the space-like region, which will allow for a determination of the
leading order hadronic contribution to the muon anomaly, in a way
completely independent from time-like dispersion relation and Lattice
QCD approaches.

The NNLO leptonic corrections consist of virtual and real
contributions. The former have been calculated with dispersion
relation techniques including all finite mass effects. For the class
of vertex corrections, the numerical results have been cross-checked
with exact analytical formulae, valid for arbitrary internal and
external masses, finding excellent agreement. Also the hadronic
contributions have been included by means of two different
parameterisations existing in the literature.  The real corrections
have been calculated by means of exact matrix elements, taking into
account all finite mass effects.

On the phenomenological side, we analysed all different,
gauge-invariant, classes of virtual contributions, investigating their
numerical impact on the differential observables relevant for MUonE
realistic running conditions. As expected, the leading contributions
come from the closed electron loop diagrams and the overall size of
the factorised and vertex contributions is of the order of few
$0.01\%$ (positive the former, negative the latter).  The vacuum
polarisation insertions on the NLO QED photonic corrections can reach
the level of few $0.1\%$, with shapes driven by the NLO photonic
corrections. The corrections are negative for all distributions,
except for the electron scattering angle distribution, where they
become positive, reaching the few $\%$ level for $\theta_e \to
0$. Similarly to what happens to the large QED NLO correction to
$d\sigma/d\theta_e$, the introduction of the acoplanarity cut
mitigates the effects and brings the corrections to negative values.
On the other hand the acoplanarity cut enhances the size of the
corrections of IR origin.

The hadronic contributions are of the same order of magnitude of the
muon loop insertions, as already noticed in previous studies on
Bhabbha scattering at flavour factories.

Particular attention has been paid to the real leptonic pair
emission. Because of the low centre of mass energy combined with
realistic event selection criteria, the $\mu^+ \mu^-$ pairs do not
play almost any role at MUonE.  On the other hand, the emission of
$e^+ e^-$ pairs is particularly important. In general the processes
$\mu^\pm e^- \to \mu^\pm e^- e^+ e^-$ are distinguishable from the
elastic $\mu^\pm e^- \to \mu^\pm e^-$ processes. However, under
realistic MUonE event selections, the $2 \to 4$ processes can produce
only two charged tracks in the final state and become, therefore,
degenerate with the elastic signal events. In this case the real pair
emission cross section has to be summed to the virtual pair emission
contribution, to reach a reliable theoretical prediction.  Because of
the presence of two indistinguishable electrons and peripheral
diagrams, we observed very large positive contributions for small
$\theta_e$ and $\theta_\mu$ as well as for small $t_{ee}$ and
$t_{\mu\mu}$ values, which are not tamed by the previously introduced
acoplanarity cut.  In order to control these effects we studied a
combination of additional cuts, namely a small minimum scattering
angle for the two observed tracks and a maximum distance from the
elasticity correlation curve $\theta_\mu(\theta_e)$. After considering
all the above cuts, the size of the real pair radiation effects has
been shown to remain below the $10^{-5}$ level.

Therefore, the combined effects of virtual and real pair radiation
under the simplified event selection we considered are dominated by
virtual contributions, in particular the ones stemming from electron
loops, which, depending on the observable, can reach the $1\%$ level
in units of the LO differential cross sections. It is true that there
is a partial cancellation between a subset of the virtual and a subset
of the real contributions, but the presence of peripheral diagrams in
real pair emission spoils the expected cancellation and forces the
introduction of elasticity cuts to keep under control a potentially
large background to the signal.

The work presented in this paper represents an additional step towards
the implementation of a fully fledged MC generator including the
complete set of NNLO QED corrections matched to multiple photon
emission, which will be ultimately needed for the analysis of MUonE
data.

\acknowledgments We are sincerely grateful to all our MUonE colleagues
for stimulating collaboration and many useful discussions, which are
the framework of the present study. We are particularly indebted to
Giovanni Abbiendi, Matteo Fael, Pierpaolo Mastrolia and Massimo
Passera for carefully reading the manuscript and for useful feedback.
One of the authors would like to thank Dr. Ulrich Schubert for useful discussions.

\appendix
\section{Master Integrals for vertex corrections}
\label{sec:Syedappendix}
\begin{figure}[ht]
        \begin{center}
          \includegraphics[width=0.4\textwidth]{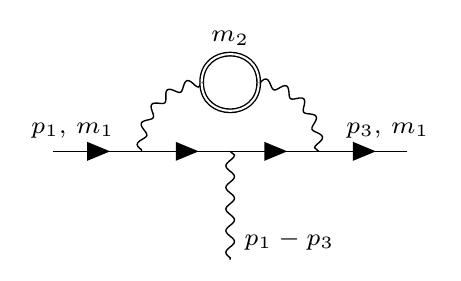}\hspace{0.5cm}
        \caption{Generic NNLO diagram with photon vacuum polarisation
          insertion in the vertex correction. The single fermionic
          line represents the leptonic current of mass $m_1$, the
          double solid line represents leptonic loop of mass $m_2$ and
          the wavy line represents photon.}
          \label{Fig:vbub}
        \end{center}
\end{figure}
The computation of the Feynman diagrams in figure~\ref{Fig:irred-vert}
is obtained by calculating the vertex form factors of the generic
diagram shown in figure~\ref{Fig:vbub}, where $m_1$ is the external
lepton mass, whereas $m_2$ represents the lepton loop mass. For the
simpler case of $m_1=m_2$, there are known results in
literature~\cite{Bonciani:2003cj,Bonciani:2003te} that can be
used. However, to the best of our knowledge, the complete analytic
expression for the case $m_1 \neq m_2$ without any expansion in the
mass ratio is not yet available. In this section we show a way to
calculate these vertex form factors for $m_1 \neq m_2$.

The kinematics of the process is given by
\begin{align}
p_1^2&=p_3^2=m_1^2 \,, & t&=(p_1-p_3)^2\,.
\end{align} 
The integrals that appear in the calculation are of the form
\begin{equation}
I(n_1,n_2,\cdots,n_7) \equiv C(\eps)\int d^dk_1 d^dk_2\frac{1}{D_1^{n_1} D_2^{n_2}\cdots D_7^{n_7}},
\end{equation}
where $C(\eps)$ is a normalisation factor:
\begin{equation}
C(\eps)=\left(\frac{m_1^2}{\mu^2}\right)^{2\eps} \frac{1}{\left[i\pi^{d/2} \Gamma(1+\eps)\right]^2}.
\end{equation}
The propagators are defined as:
\begin{align*}
D_1 &= k_1^2,\\
D_2 &= k_2^2 -m_2^2,\\
D_3 &= (k_1+k_2)^2 -m_2^2,\\
D_4 &= (k_1-p_1)^2 -m_1^2, \\
D_5 &= (k_1-p_3^2)^2-m_1^2,\\
D_6 &= (k_1+k_2-p_1)^2,\\
D_7 &= (k_1+k_2-p_3)^2.
\end{align*}

\begin{figure}
\centering
\captionsetup[subfigure]{labelformat=empty}
\subfloat[$\mathcal{T}_1$]{
    \includegraphics[width=0.2\textwidth]{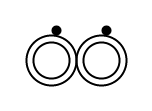}
  } \,
  \subfloat[$\mathcal{T}_2$]{
    \includegraphics[width=0.2\textwidth]{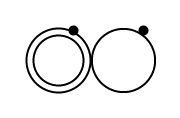}
  }\,
  \subfloat[$\mathcal{T}_3$]{
    \includegraphics[width=0.2\textwidth]{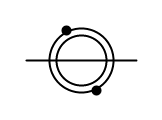}
  }\,
  \subfloat[$\mathcal{T}_4$]{
    \includegraphics[width=0.2\textwidth]{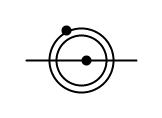}
  }\,
  \subfloat[$\mathcal{T}_5$]{
    \includegraphics[width=0.25\textwidth]{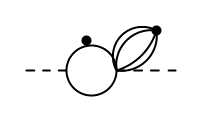}
  }
  \subfloat[$\mathcal{T}_6$]{
    \includegraphics[width=0.25\textwidth]{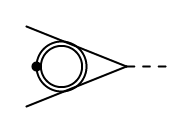}
  }
  \subfloat[$\mathcal{T}_7$]{
    \includegraphics[width=0.25\textwidth]{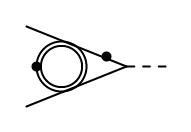}
  }\,
 \caption{The master integral topologies $\mathcal{T}_{1\dots7}$ for
   fermionic loop induced NNLO vertex correction.  The solid single
   line represents a propagator with mass $m_1$ and the double line
   represents a propagator with mass $m_2$, while the dashed line
   represents a massless propagator with momentum squared equal to
   $t$. The dots represent additional powers of the propagator.  }
 \label{fig:MIs}
\end{figure}

Seven master integrals, depicted in figure~\ref{fig:MIs}, are
necessary to calculate the form factors. We solved these particular
integrals using differential equation
methods~\cite{Kotikov:1990kg,Remiddi:1997ny,Gehrmann:1999as}: we chose
a suitable basis of master integrals, such that the dimensional
regularisation parameter $\eps$ factorises from the kinematics. The
integrals were then encoded in a $\dlog$-form, also called canonical
form~\cite{Henn:2013pwa}, which we were able to obtain using the
Magnus algorithm~\cite{DiVita:2014pza,Argeri:2014qva}. The kinematic
variables were chosen in such a way that the arguments of the
$\dlog$'s were simple rational functions, which enabled us to express
the integrals in terms of the generalised
polylogarithms~\cite{Remiddi:1999ew,Gehrmann:2001jv,Vollinga:2004sn}. The
master integrals were then calculated up to required expansion to
determine the form factor: they are $I(0,2,2,0,0,0,0)$, $I(0, 2, 0, 2,
0, 0, 0)$, $I(0, 2, 2, 1, 0, 0, 0)$,
$I(0, 2, 1, 2, 0, 0, 0)$, $I(0, 2, 0, 2, 1, 0, 0)$,
$I(0, 1, 2, 1, 1, 0, 0)$ and $I(0, 1, 2, 2, 1, 0, 0)$,
which can be found in the ancillary file in the \texttt{arXiv}
version of the paper. Here we explicitly write the first four master
integrals up to finite term to show the used convention.

\begin{align}
I(0,2,2,0,0,0,0)&= \frac{1}{\epsilon ^2}+\frac{4G(\{0\},z)}{\epsilon}
+16 G(\{0,0\},z)+O\left(\epsilon ^1\right)\\ I(0, 2, 0, 2, 0, 0, 0)&
=\frac{1}{\epsilon ^2}+\frac{2G(\{0\},z)}{\epsilon }+4
G(\{0,0\},z)+O\left(\epsilon ^1\right)\\ I(0, 2, 2, 1, 0, 0,
0)&=\frac{(z+2) G(\{-1,0\},z)}{m_1^2}+\frac{(2-z)
  G(\{1,0\},z)}{m_1^2}+O\left(\epsilon ^1\right)\\ I(0, 2, 1, 2, 0, 0,
0)&
=-\frac{G(\{-1,0\},z)}{m_1^2}-\frac{G(\{1,0\},z)}{m_1^2}+O\left(\epsilon
^1\right)
\end{align}

We introduced the dimensionless variables $x$ and $z$ such that the
arguments of the $\dlog$ were simple rational functions:

\begin{align}
    t&=-\frac{m_1^2 \left(x^2-2 x z-3 z^2+4\right)^2}{2
      \left(x^2-z^2\right) \left(x z+z^2-2\right)}, &
    m_2^2&=\frac{m_1^2}{z^2}.
\end{align}
The following equations show the solutions that are suitable for the
region of interest:

\begin{equation}
    \begin{split}
        x&= \!\begin{multlined}[t][13 cm] -\frac{1}{2 m_1
          m_2}\Bigg[-\sqrt{4 m_1^2-t} \sqrt{4 m_1^2-4 m_2^2-t}\\ +
          \sqrt{2} \sqrt{-t \left(\sqrt{4 m_1^2-t} \sqrt{4 m_1^2-4
              m_2^2-t}+4 m_1^2-2 m_2^2-t\right)}-2 m_1^2+t \Bigg],
        \end{multlined}\\   
        z& =\frac{m_1}{m_2}
    \end{split}
\end{equation}

Throughout the form factor calculation, the Integration by Parts
reduction ~\cite{Tkachov:1981wb,Chetyrkin:1981qh,Laporta:2001dd} was
carried out using the publicly available codes
\textsc{Reduze}~\cite{vonManteuffel:2012np} and
\textsc{FIRE}~\cite{Smirnov:2019qkx}. The differential equations were
generated using \textsc{Reduze} and, to rationalise the arguments of
the $\dlog$, we used the package
\textsc{RationalizeRoots}~\cite{Besier:2019kco}. The numerical
validations were accomplished using
\textsc{SecDec}~\cite{Borowka:2015mxa} and the numerical evaluation of
generalised polylogs was performed with
\textsc{GiNaC}~\cite{Vollinga:2004sn} and
\textsc{handyG}~\cite{Naterop:2019xaf}. A detailed result of the
vertex form factor is presented in ref.~\cite{Syed:2021nc}.

The resulting expressions for the form factors have been used to
cross-check with high accuracy the DR results of
figures~\ref{Fig:vertex-blob-on-thetae-thetamu}
and~\ref{Fig:vertex-blob-on-t}.

\bibliographystyle{JHEP}
\bibliography{muone_pairs}
\end{document}